\documentclass{pasa}%
\usepackage{color}

\title[Sampling of spectra]{Detector sampling of
  optical/IR spectra: how many pixels per FWHM?}
\author[J.G. Robertson]{J. Gordon
  Robertson$^{1,2}$\thanks{G.Robertson@physics.usyd.edu.au}\\   
\affil{$^1$Sydney Institute for Astronomy, School of Physics, University of
  Sydney, NSW 2006, Australia}% 
\affil{$^2$Australian Astronomical Observatory, PO Box 915, North Ryde, NSW
  1670, Australia}}% 
\jid{PASA}
\doi{10.1017/pas.\the\year.xxx}
\jyear{\the\year}

\begin{document}%
\begin{abstract}
  Most optical and IR spectra are now acquired using detectors with finite-width pixels in a
  square array. Each pixel records the received intensity integrated over its own area, and
  pixels are separated by the array pitch. This paper examines the effects of such pixellation,
  using computed simulations to illustrate the effects which most concern the astronomer
  end-user. It is shown that coarse sampling increases the random noise errors in wavelength by
  typically $10 - 20$ \% at 2 pixels/FWHM, but with wide variation depending on the functional
  form of the instrumental Line Spread Function (LSF; {\it i.e.} the instrumental response to a
  monochromatic input) and on the pixel phase. If line widths are determined they are even more
  strongly affected at low sampling frequencies. However, the noise in fitted peak amplitudes
  is minimally affected by pixellation, with increases less than about 5\%.  Pixellation has a
  substantial but complex effect on the ability to see a relative minimum between two
  closely-spaced peaks (or relative maximum between two absorption lines).  The consistent
  scale of resolving power presented by Robertson (2013) to overcome the inadequacy of the Full
  Width at Half Maximum (FWHM) as a resolution measure is here extended to cover pixellated
  spectra. The systematic bias errors in wavelength introduced by pixellation, independent of
  signal/noise ratio, are examined. While they may be negligible for smooth well-sampled
  symmetric LSFs, they are very sensitive to asymmetry and high spatial frequency
  substructure. The Modulation Transfer Function for sampled data is shown to give a useful
  indication of the extent of improperly sampled signal in an LSF. The common maxim that 2
  pixels/FWHM is the Nyquist limit is incorrect and most LSFs will exhibit some aliasing at
  this sample frequency. While 2 pixels/FWHM is nevertheless often an acceptable minimum for
  moderate signal/noise work, it is preferable to carry out simulations for any actual or
  proposed LSF to find the effects of various sampling frequencies. Where spectrograph
  end-users have a choice of sampling frequencies, through on-chip binning and/or spectrograph
  configurations, it is desirable that the instrument user manual should include an examination
  of the effects of the various choices.
\end{abstract}
\maketitle %
\section{INTRODUCTION}
\label{sec:introduction}
Detectors with pixels on a square array are in widespread use in current optical and IR grating
spectrographs. A key issue confronting the designer or user\footnote{Users may have to decide
  on the slit width or grating configuration and whether to use on-chip binning, all of which
  can affect the sampling rate.} of such instruments is the size of the individual detector
pixels with respect to the width of the instrumental profile - in other words how many pixels
should be used to sample the width of an unresolved spectral line. If too few pixels are used,
narrow or unresolved spectral features will be undersampled, resulting in errors of position
({\it i.e.} wavelength) and possibly also flux which vary depending on the phase of the
spectral line with respect to the pixel centres. The ability to distinguish closely spaced
lines will be compromised, as will detection of slight broadening of a spectral feature. Random
errors in wavelength and width are also increased with coarse sampling. On the other hand, if an
unnecessarily large number of pixels are used to span the instrumental width, then the
spectrograph's total wavelength range will be reduced for a given detector size, and the
effects of readout noise, dark noise and cosmic ray hits will be exacerbated.
 
The approach that has most often been taken in the literature is to say that 2 pixels per Full
Width at Half Maximum (FWHM) represents Nyquist sampling, which is satisfactory but a bare
minimum, and that for more accurate work (such as accurate radial velocities or velocity
dispersions) a larger number of pixels such as 3 to 4 per FWHM should be used.  However, 2
pixels per FWHM is not the Nyquist limit and the minimum acceptable sampling frequency
depends on the required accuracy and on the form of the instrumental profile, here referred to
as the Line Spread Function (LSF).

The CCD and IR array detectors perform two operations on the LSF as it falls on the detector,
in the process of sampling ({\it e.g.} Bernstein 2002): firstly the light which falls within
the area of a single pixel is integrated over the finite area of the pixel, and secondly a
single intensity value is recorded and nominally ascribed to the location at the centre of the
pixel. In this work the pixels will be regarded as having uniform sensitivity across their
width (except in Section \ref{sec:non-uniform}).

Using Fourier Transform methods it is quite straightforward to take any proposed LSF, convolve
it with a rectangle representing the smoothing effect of integrating the signal over the pixel
width, and then find the extent to which the Fourier components lie beyond the true Nyquist
frequency ({\it i.e.} 2 samples per cycle of a sinusoid) for any given sampling
frequency. However, this does not answer the questions which concern the instrument scientist
or astronomer, namely what level of errors will be introduced into the measurement of
wavelengths, line strengths and widths, and how will the ability to separate two closely-spaced
spectral lines be affected?  This paper aims to make a start towards answering these more
practical questions. The approach will be to illustrate the effects of sampling, rather than to
give another mathematical analysis.

Spectra will be assumed to be 1-dimensional, {\it i.e.} representing an array giving intensity as a
function of wavelength. The integration or extraction over the spatial dimension to produce
such 1-dimensional spectra is not the issue here, and the emphasis is on adequacy or otherwise
of sampling as a function of wavelength. Data samples will be assumed to be digitised with
sufficient precision that quantisation noise can be neglected. 

A comprehensive introduction to the basics of the sampling process is given by Vollmerhausen et
al (2010); see also Anderson \& King (2000). Bickerton \& Lupton (2013) presented Fourier
methods to give accurate photometry of sampled images.

\section{EFFECTS OF SAMPLING} 
\label{sec:effect of sampling}
Sampling by pixels with uniform sensitivity and no interpixel gaps (as assumed above) has a
number of effects:

1) The integration of the received intensity signal over the width of a pixel has the effect of
smoothing the incident intensity profile, through convolution with a rectangular profile having
the width of one pixel. This broadens the effective LSF and hence reduces the spectral
resolving power of the instrument, in the sense that it cannot resolve two closely spaced
spectral lines as well as before sampling.

2) The noise errors of the key parameters of spectral lines - namely position (wavelength)
 and width will be increased, and the flux errors will be increased unless an integrated flux
 is used.

3) If sampling is inadequate ({\it i.e.} too coarse) then systematic bias errors may be introduced in
the fitted line parameters.

4) Coarse sampling may reduce the ability to distinguish closely spaced spectral features.

While simple in principle, the quantification of the above errors is complicated by the
fact that the errors do not depend solely on the sampling rate and the LSF functional form -
they also depend on the type of analysis performed on the spectral data. This analysis can take
many forms - such as fitting a Gaussian or series of Gaussians to spectral features (although the
LSF will in general not be an exact Gaussian), or fitting another functional form, or
calculation of a centroid wavelength, or perhaps interpolation (referred to as `reconstruction'
in the literature on 2D imaging). An important special case is the fitting of the exactly
correct LSF (including allowance for the convolution effect of the finite pixel width) to unresolved
spectral features. In this case unbiased wavelengths can be obtained even with significant
undersampling, because the process is analogous to deconvolution. 

The analysis and discussion below attempts to find within this complex multi-dimensional
problem some results which are useful to designers and users of sampled spectrographs.

\section{WAVELENGTH ACCURACY} 
\label{sec:wavelength accuracy}

It is expected that coarse pixellation will increase the random wavelength uncertainty in
locating an unresolved spectral feature. The conceptually simplest view is that the effective
LSF is the intrinsic (instrumental) LSF convolved with the pixel rectangle - this necessarily
broadens the LSF and hence diminishes the wavelength accuracy. However, the simple view does
not take into account the dependence on pixel phase, which is important at low sampling
frequency, so we proceed as follows. (The results here are given assuming positive-going peaks,
but would be equally applicable to weak absorption features which leave the per-pixel noise
approximately constant.)

Defining $\sigma$ as the rms noise in each pixel (wavelength channel) and assuming the noise is
constant per unit wavelength interval (at least in the vicinity of a given spectral feature)
and that the noise in separate pixels is uncorrelated, the formula for wavelength uncertainty
$\sigma_{\lambda}$ given by Clarke et al. (1969) can be used\footnote{See Robertson (2013)
  regarding correction of the typographical error in equation (A7) of Clarke et al. (1969). This
  is the wavelength uncertainty for a two-parameter fit (peak amplitude and wavelength); width
  is assumed unresolved and is not fitted. Fitting the exact LSF with the minimum number of
  parameters is the optimum process (as compared with fitting a Gaussian to a non-Gaussian LSF,
  or using the centroid etc) so the resulting $\sigma_{\lambda}$ will be the minimum
  achievable. Note also that what is evaluated here is the signal/noise - dependent random
  error, not the bias error, which is considered in Section \ref{sec:bias}.}:

\begin{equation}
\sigma_{\lambda}^2 = \frac{\sigma^2}{ {\rm pk}^2 \Big( \sum{(B')^2} - \Big(\sum{BB'}\Big)^2 /
                   \sum{B^2} \Big)}
\label{eqn:sig_x_full}
\end{equation}

where `pk' is the peak amplitude of the response whose $\sigma_{\lambda}$ is to be found.
% what is 'pk' for absorption lines??
$B'$ is the derivative with respect to wavelength of the LSF $B$, which is assumed normalised
to a peak of unity (before pixel convolution). Some care is needed in the evaluation of $B'$
when pixels are wide and the slope may change greatly across one pixel. Inspection of the
Clarke et al. derivation shows that $B'$ for a given pixel refers to $dB/d\lambda$ where $B$ is
interpreted as the continuous ({\it i.e.} unsampled) LSF but integrated across the pixel and
then point sampled at the relevant pixel centre. Thus eqn (\ref{eqn:sig_x_full}) may be
conveniently evaluated for any LSF that is known as a continuous function, by first convolving
with the pixel rectangle and then sampling $B$ and $B'$ at the appropriate locations. The
validity of the formula for coarsely pixellated data treated in this way has been verified by
Monte Carlo tests.

In the case of a well-sampled symmetric LSF observing an unresolved spectral feature, this
equation simplifies to :

\begin{equation}
\sigma_{\lambda}^2 = \sigma^2 / \Big( {\rm pk}^2 \sum{(B')^2}\Big).
\label{eqn:sig_x_simple}
\end{equation}

However, the general form (\ref{eqn:sig_x_full}) will be used here in order to handle coarse
sampling correctly. The results for a number of LSF forms will now be examined.

\begin{figure}[h]
\begin{center}
\includegraphics*[angle=0, scale=0.417]{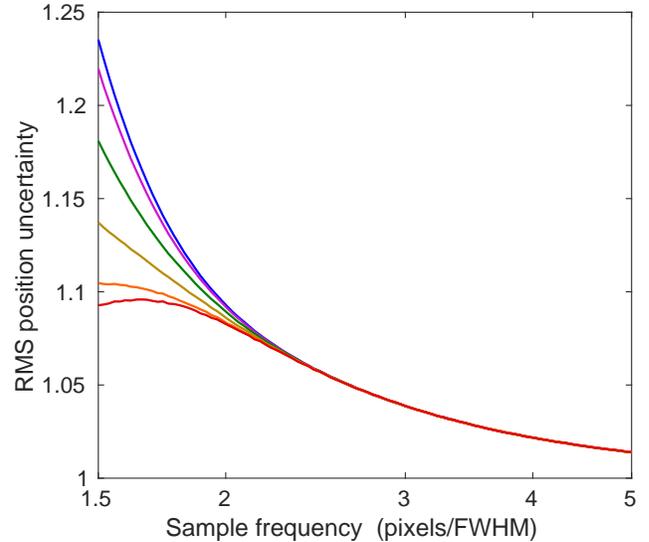} % was 0.45, 0.433
\caption{Wavelength uncertainty vs sampling frequency for a range of pixel phase values. The
  LSF form is Gaussian. The uppermost (blue) curve corresponds to pixel phase = 0 (peak centred
  on a pixel) while the lowest (red) curve, which has a maximum at sample frequency $\sim
  1.64$, is for pixel phase = $\pm0.5$, {\it i.e.} a peak lying on the boundary between two
  pixels. Other curves have pixel phases at intervals of 0.1. The noise in each pixel is
  uncorrelated and is appropriately scaled for the actual pixel widths, such that the noise for
  unit dispersion axis interval remains constant. The vertical scale of RMS position uncertainties has
  been normalised to unity at very large sample frequency.}
\label{fig:Fig_7_23_a}
\end{center}
\end{figure}

\begin{figure}[h]
\begin{center}
\includegraphics[angle=0, scale=0.3]{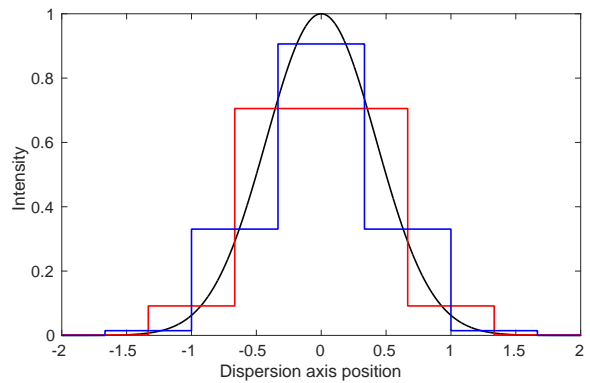}
\caption{Gaussian LSF (black) with superposed sampled versions of the same LSF (blue: pixel
  phase 0, red: pixel phase 0.5). The sampling frequency is 1.5 pixels/FWHM.}
\label{fig:Fig_6_53_gauss_b} 
\end{center}
\end{figure}

Figure \ref{fig:Fig_7_23_a} shows the variation of random wavelength error with
sampling frequency for the case of a Gaussian LSF. All curves show a rise in position errors as
the sampling frequency decreases.  This is expected due to the lessened sensitivity of wide
pixels to the steep sides of the LSF. Below about 2 pixels per FWHM there is increasing
dependence of the result on the pixel phase. The greatest errors occur for the LSF peak centred
in the middle of a pixel, while for the LSF centred on the boundary between two pixels, errors
are minimised and there is actually a turn-over, with coarser pixellation resulting in lower
position errors. This can be understood as the pair of pixels starting to act as one axis of a
quad-cell position locator ({\it i.e} using the flux ratio between the two adjacent pixels to
determine the peak location).  The red line in Figure \ref{fig:Fig_6_53_gauss_b} shows the LSF
straddling the boundary of two pixels for a sampling frequency of 1.5 pixels/FWHM, illustrating
the rapid shift of flux between the two major pixels, as the peak shifts slightly with respect
to the pixels.

\begin{figure}[h]
\begin{center}
\includegraphics*[angle=0, scale=0.413]{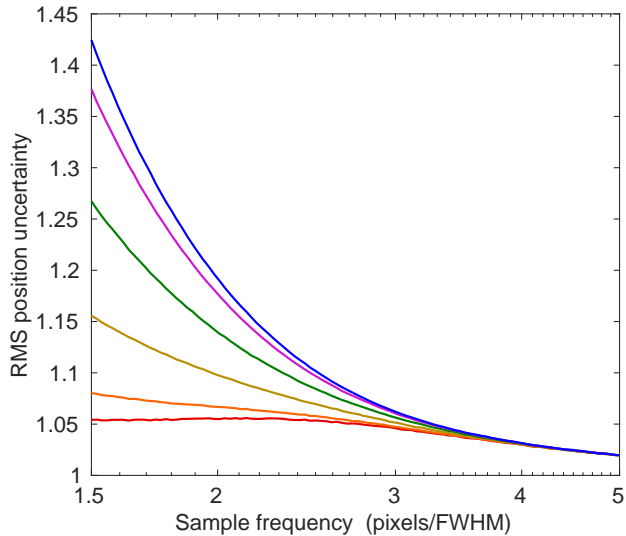} % was 0.455,0.438 
\caption{Wavelength uncertainty vs sampling frequency for a Lorentzian LSF, otherwise as for
  Figure \ref{fig:Fig_7_23_a}.}\label{fig:Fig_7_23_b}
\end{center}
\end{figure}

\begin{figure}[h]
\begin{center}
\includegraphics[angle=0, scale=0.3]{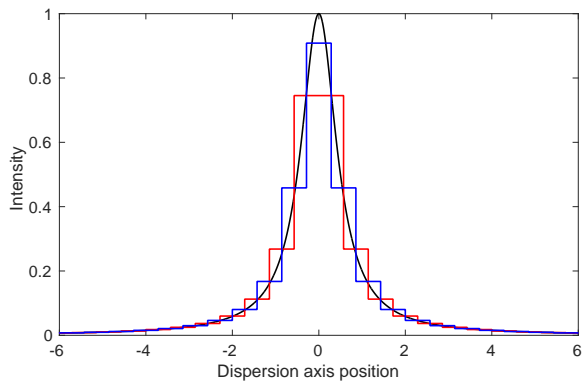}
\caption{Lorentzian LSF (black) with superposed sampled versions of the same LSF (blue: pixel
  phase 0, red: pixel phase 0.5).  The sampling frequency is 1.75 pixels/FWHM.}
\label{fig:Fig_6_54_lorentz_b} 
\end{center}
\end{figure}

Figure \ref{fig:Fig_7_23_b} shows the similar set of curves, in this case for a Lorentzian
LSF. The sharpness of the Lorentzian's core with respect to its wings results in a pronounced
dependence of the wavelength error on pixel phase. Thus the red line in Figure
\ref{fig:Fig_6_54_lorentz_b}, at pixel phase 0.5, shows the case where the `quad-cell' effect
holds the wavelength error nearly constant with decreasing sample frequency. On the other hand,
the blue line, at pixel phase 0, corresponds to sharply increased $\sigma_{\lambda}$.

\begin{figure}[h]
\begin{center}
\includegraphics*[angle=0, scale=0.41]{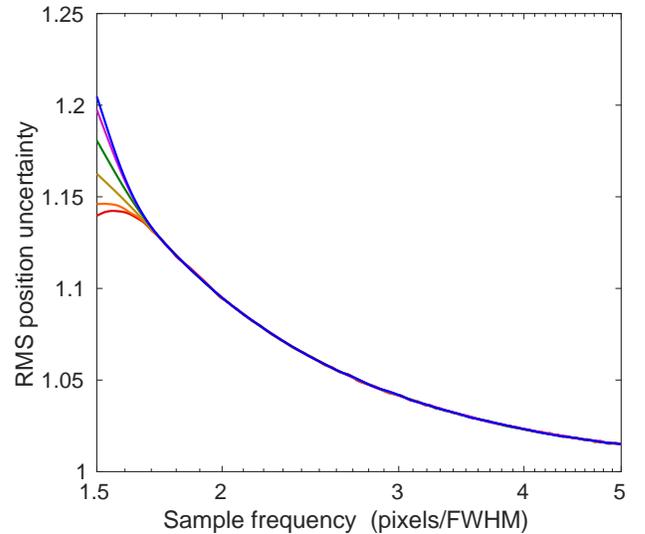} % was 0.45, 0.433
\caption{Wavelength uncertainty vs sampling frequency for a sinc$^2$ LSF. Since the sinc$^2$
  function has minor lobes with amplitudes decaying slowly away from the central peak, it is
  not possible to include all the function's non-zero values as was effectively done for other
  LSFs. In this case the summations were continued to dispersion axis positions of $\pm 225.76
  \times {\rm FWHM}$, in order to obtain a close approximation to the band-limited nature of
  sinc$^2$. The pixel phase curves are coloured as in Figure \ref{fig:Fig_7_23_a}.}
\label{fig:Fig_7_23_c}
\end{center}
\end{figure}

\begin{figure}[h]
\begin{center}
\includegraphics*[angle=0, scale=0.41]{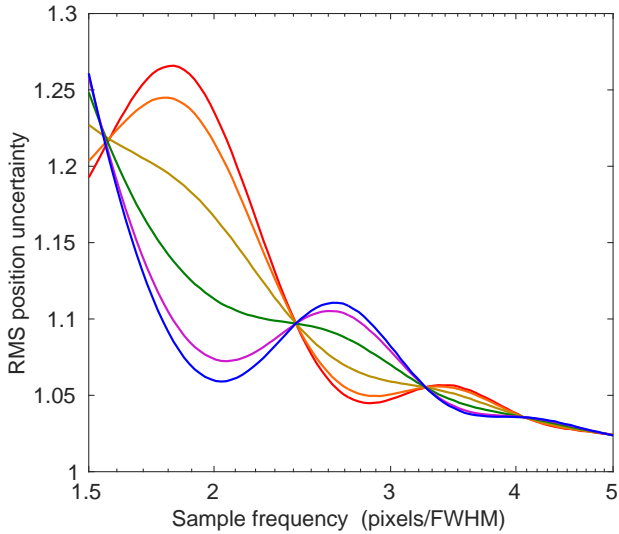} % was 0.45,0.433
\caption{Wavelength uncertainty vs sampling frequency for a range of pixel phase values. The
  LSF form is the convolved projected circle. At a sample frequency of 2 pixels/FWHM, the {\it
    lowest} (blue) curve corresponds to pixel phase = 0 (peak centred on a pixel) while the
  highest (red) curve is for pixel phase = $\pm0.5$.  Other curves again have pixel phases at
  intervals of 0.1.}
\label{fig:projcircleconv_sigma_lambda}
\end{center}
\end{figure}

\begin{figure}[h]
\begin{center}
\includegraphics[angle=0, scale=0.3]{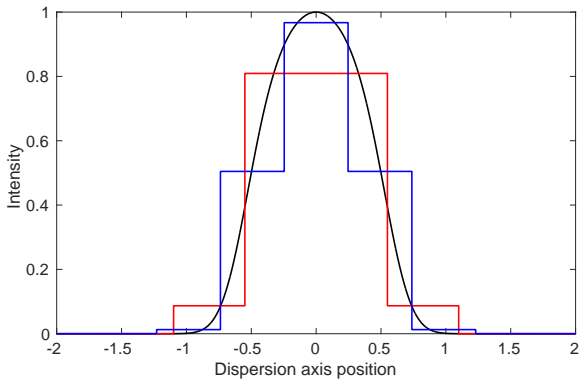}
\caption{Convolved projected circle LSF (black) with superposed sampled versions of the same LSF
  (blue: pixel phase 0 at 2.03 pixels/FWHM ({\it i.e.} at the local minimum of the position
  uncertainty curve), red: pixel phase 0.5 at sampling frequency 1.82 pixels/FWHM ({\it i.e.}
  at the local maximum)).}
\label{fig:Fig_6_54_prjcrcconv_b}        
\end{center}
\end{figure}

\begin{figure}[h]
\begin{center}
\includegraphics*[angle=0,  scale=0.41]{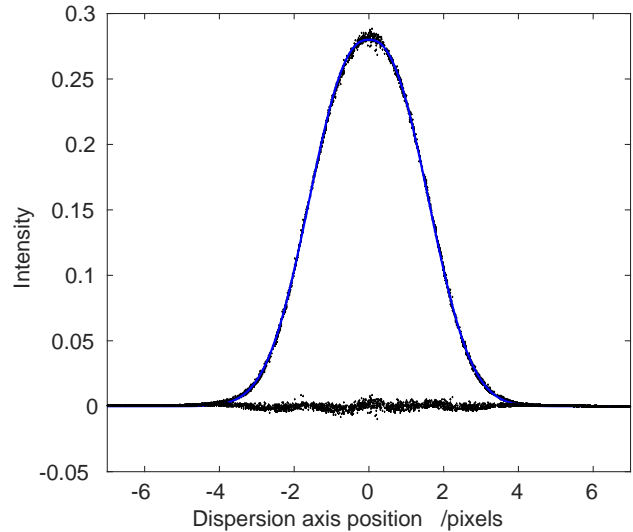} % was 0.35
\caption{Arc line profile from AAOmega, using a subarea of 1000 (spatial) x 500 (wavelength)
  pixels near the centre of the data frame. 
 %subarea x (spatial) = 1250 - 2250 and y
 % (wavelength) = 750 - 1250 {\it i.e.} staying away from the edges of the data frame.  
  266 unblended and unsaturated fibre images were selected for processing. The horizontal axis
  is in units of pixels, and the vertical axis is intensity in arbitrary units.  The blue curve
  is a 3-parameter empirical fit, $I = 0.2800 \exp(-0.1854|p|^{2.4174})$ where $p$ is the
  horizontal axis independent parameter in pixels. The residuals with respect to this fit are
  also shown - indicating that the fit is good but not perfect. The fitted curve has FWHM =
  3.450 pixels.}\label{fig:Fig_7_23_e}
\end{center}
\end{figure}

\begin{figure}[h]
\begin{center}
\includegraphics*[angle=0, scale=0.41]{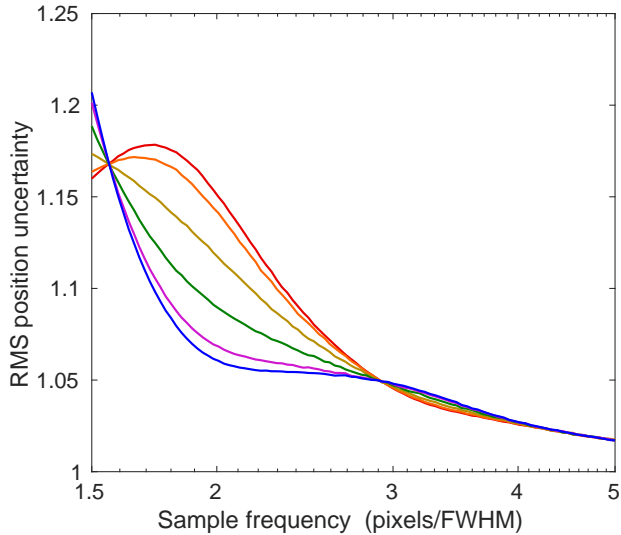} % was 0.45,0.433
\caption{Wavelength uncertainty vs sampling frequency for a range of pixel phase values. The
  LSF is from the AAOmega spectrograph.  At a sample frequency of 2 pixels/FWHM, the {\it
    lowest} (blue) curve corresponds to pixel phase = 0 (peak centred on a pixel) while the
  highest (red) curve is for pixel phase = $\pm0.5$.  Other curves again have pixel phases at
  intervals of 0.1. (The actual sample frequency for the intrinsic LSF of AAOmega is 3.41
  pixels/FWHM.)}
\label{fig:AAOmega_sigma_lambda}
\end{center}
\end{figure}

\begin{figure}[h]
\begin{center}
\includegraphics[angle=0,scale=0.3]{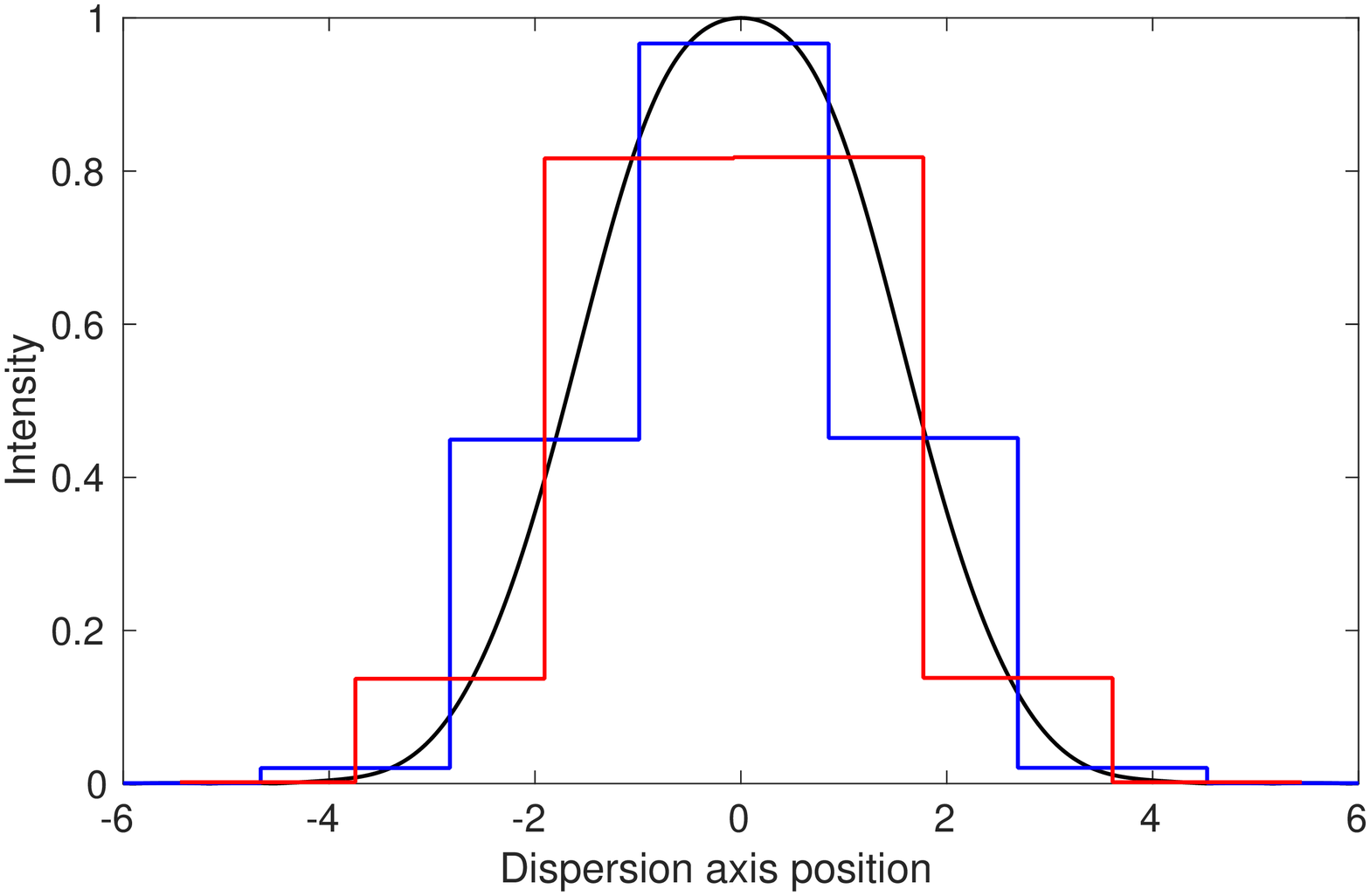} 
\caption{AAOmega spectrograph LSF (black) with superposed sampled versions of the same LSF
  (blue: pixel phase 0, red: pixel phase 0.5).  The sampling frequency is 2.0 pixels/FWHM
  ({\it not} equal to the actual as-built pixel scale). The unsampled LSF has a FWHM of 3.41.}
\label{fig:Fig_6_54_aaomega_b} 
\end{center}
\end{figure}

Figure \ref{fig:Fig_7_23_c} shows the  $\sigma_{\lambda}$ vs sampling frequency curves for a
sinc$^2$ LSF. This LSF form arises in the case of a diffraction-limited slit, and has the
important property that its Fourier Transform is band-limited, {\it i.e.} if it is sampled at
or above the Nyquist rate, there will be no aliasing. This topic will be expanded in Section
\ref{sec:fourier} but the important point to note from Figure \ref{fig:Fig_7_23_c} is that the
divergence of the curves for different pixel phases begins below the sampling frequency of 1.77
pixels/FWHM, which corresponds closely to Nyquist sampling. In other words, where there is no
aliasing, there is also no dependence of  $\sigma_{\lambda}$ on pixel phase.

The next LSF to be considered is intended to represent the result of projecting the image of a
multi-mode fibre on to the wavelength axis. The ideal result is a half ellipse ({\it e.g.} Bracewell
1995), but the extremely steep sides of the ellipse will inevitably be smoothed to some extent
by optical aberrations in the spectrograph, so what is used here is the projected circle
convolved with the Gaussian that produces the minimum final FWHM (Robertson 2013, hereafter
referred to as Paper 1). This provides an example representative of projected multi-mode fibres
subject to some spectrograph aberrations. Figure \ref{fig:projcircleconv_sigma_lambda} shows
the variation of random wavelength errors with sampling frequency and pixel phase. The
oscillatory behaviour, with pixel phase 0 being best at some sampling frequencies, and phase
0.5 best at others, is radically different from the Gaussian, Lorentzian and sinc$^2$
cases. The reason is the steep-sided and flat-topped nature of the LSF. As equations
\ref{eqn:sig_x_full} and \ref{eqn:sig_x_simple} show, the accuracy of wavelength determination
depends on the regions of greatest slope. Thus $\sigma_{\lambda}$ will be minimised at the
pixel phases and sampling frequencies that are able to derive good position discrimination from
both sides of the LSF. The red line in Figure \ref{fig:Fig_6_54_prjcrcconv_b} shows the sampled
LSF at pixel phase 0.5 and sample frequency 1.82 pixels/FWHM, which is a local maximum in
$\sigma_{\lambda}$. In this configuration there is minimal sensitivity of the relative pixel
intensities to small changes in wavelength, due to the flat-topped nature of the LSF. On the
other hand the blue line in Figure \ref{fig:Fig_6_54_prjcrcconv_b} shows pixel phase 0 at 2.03
pixels/FWHM, giving a local minimum of $\sigma_{\lambda}$. Here, a small shift of the LSF with
respect to the pixels produces a maximal change in values for the pixels on either side of the
centre. The flattened top of the LSF in effect decouples the contributions of the two sides,
resulting in particular combinations of pixel phase and sample frequency that are optimum.

The final LSF is that of an actual spectrograph, the AAOmega instrument operating at the
Anglo-Australian Telescope (Saunders et al. 2004). This instrument uses multi-mode fibres and
thus can be expected to give results similar to those of the convolved projected circle. It is
nevertheless interesting to see whether curves such as Figure
\ref{fig:projcircleconv_sigma_lambda} are borne out in practice. An arc frame of CCD data from
AAOmega was used, and 266 fibre images were selected from the central part of the CCD where the
LSF is sufficiently constant. Each fibre image was projected to the wavelength axis, then
images were aligned to their centroid and scaled to flux equality and plotted together, as
shown by the black points in Figure \ref{fig:Fig_7_23_e}.  Because there is some tilt of the
lines along the AAOmega spatial axis, and also a number of lines of different wavelength were
used, the result is good coverage of all pixel phases ({\it i.e.} position of centroid with
respect to the pixel boundaries).  There are 266 fibre profiles with an average of
14.3 points per profile, giving 3800 points, hence virtually continuous coverage
of the LSF.

What this shows (as the set of black plotted points) is the true LSF received by the detector,
after convolution with the pixel response. Anderson \& King (2000) refer to this as the
effective PSF (ePSF).

As expected, this LSF cannot be satisfactorily fitted with a Gaussian, because the peak is too
broad and the wings too low relative to the best-fit Gaussian. The blue curve in Figure
\ref{fig:Fig_7_23_e} shows an empirical fit that is adequate for the present purposes. Using
that fit the LSF profile was successfully deconvolved to remove the effect of pixel
convolution, assuming the pixel response is a perfect rectangle of width 1 pixel. Because the
profile is well sampled, pixel deconvolution makes only a minimal difference.

The deconvolved LSF represents the LSF as it fell on the detector, so it can be used as input
to calculation of $\sigma_{\lambda}$ vs sampling frequency as before. Figure
\ref{fig:AAOmega_sigma_lambda} shows the results. Indeed the oscillations do occur, although at the
actual AAOmega sampling frequency of 3.41 pixels/FWHM, there is minimal dependence on pixel
phase. But if 2-pixel sampling had been adopted, it would have been quite significant.

Figure \ref{fig:Fig_6_54_aaomega_b} shows the deconvolved AAOmega LSF and its pixellation at phases
0 and 0.5.

The results given in this section show that sampling at 2 pixels/FWHM causes a loss of
typically $10 - 20$\% in wavelength accuracy relative to the limit of continuous sampling. There is,
however, a considerable variation among the different LSF forms, and increasing dependence on
pixel phase for coarse sampling.

\section{WIDTH ACCURACY} 
\label{sec:width accuracy}

Pixellation has an even greater effect on the accuracy of width measurements for barely
resolved spectral features, such as in the measurement of galaxy velocity dispersions. It is
intuitively obvious that coarse pixellation will impede the determination of an accurate width
measurement. In this section examples of two LSFs are given to illustrate this point.

When three parameters (location, peak height and width) are to be fitted to a profile, it is
not possible to give an explicit equation for the width uncertainties that is analogous to
equation (\ref{eqn:sig_x_full}). Instead the procedure adopted was to use the non-linear least
squares fitting facility {\it nlinfit} in {\sc matlab}\footnote{www.mathworks.com.au}. The
input data set was the (unbroadened) LSF sampled at the appropriate sampling frequency and
pixel phase, and the fitting function was the same LSF. With starting values for the location,
peak height and FWHM deliberately differing from the correct values by $ 5 - 10$\%, {\it
  nlinfit} then determined the best fit parameter values (using the correct integration across
pixels for the fitting function). Instead of injecting noise and then performing numerous Monte
Carlo simulations to assess the scatter of fitted parameter values, use was made of the
Jacobian matrix $J$ of the non-linear regression model, which can be returned by {\it
  nlinfit}. Calculation of $\sigma^2 (J^T * J)^{-1}$ gives the covariance matrix of the three
parameters, when the noise is assumed to be independent and have the same RMS value $\sigma$ in
each pixel.

\begin{figure}[h]
\begin{center}
\includegraphics*[angle=0, scale=0.41]{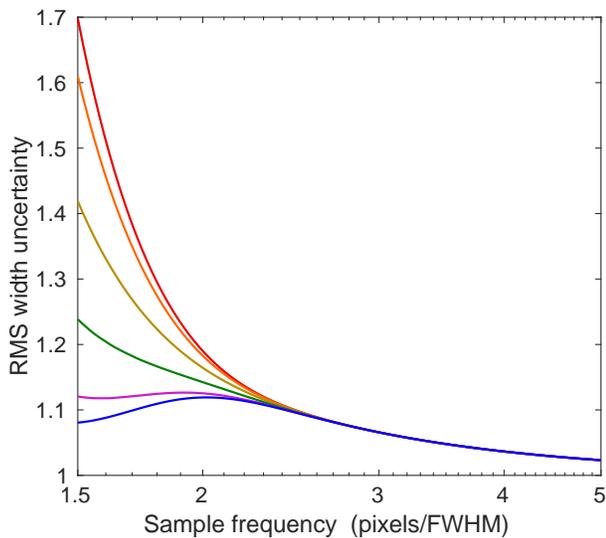}  % was 0.479
\caption{Width uncertainty vs sampling frequency for a range of pixel phase values. The LSF is
  Gaussian.  The {\it lowest} (blue) curve corresponds to pixel phase = 0 (peak centred on a
  pixel) while the highest (red) curve is for pixel phase = $\pm0.5$.  Other curves have
  pixel phases at intervals of 0.1.  The noise in each pixel is uncorrelated and is
  appropriately scaled for the actual pixel widths, such that the noise for unit dispersion
  axis interval remains constant. The vertical scale of RMS width uncertainties has been
  normalised to unity at very large sample frequency.}
\label{fig:Fig_7_23_g}
\end{center}
\end{figure}

Figure \ref{fig:Fig_7_23_g} shows the results for a Gaussian LSF. Since the input function and
the fitting function were both unbroadened, the results represent the noise in width for small
width extensions. The plots have been normalised to unity at large sample frequencies, {\it
  i.e.} the numerical values again show the factor by which sampling increases the noise. This
Figure shows that pixel phase dependence develops below about 2.4 pixels/FWHM, and worsens
rapidly below 2 pixels/FWHM. In contrast to the case for wavelength uncertainties, the lowest
errors for width are obtained at pixel phase 0 (peak centred on a pixel) and the worst at pixel
phase $\pm 0.5$ (peak at the boundary of two pixels). This can be understood since in the
latter case the width determination will rely on the values in the two pixels either side of
the central two, and there is little signal in them at low sample frequencies ({\it e.g.} the
red histogram pixels at $\pm 1$ in Figure \ref{fig:Fig_6_53_gauss_b}). Saunders (2014) also
noted that the best pixel phase for position (wavelength) determination is the worst for
widths.

\begin{figure}[h]
\begin{center}
\includegraphics*[angle=0, scale=0.41]{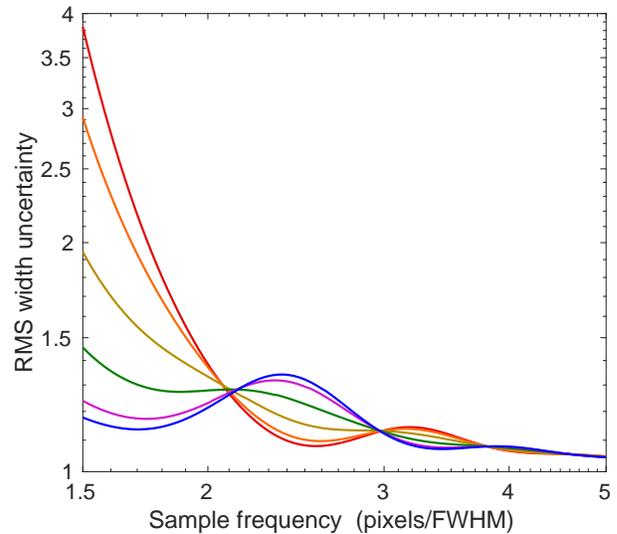} 
\caption{Width uncertainty vs sampling frequency for the convolved projected circle LSF. The
  colour coding for pixel phases is as before. }
\label{fig:Fig_7_23_h}
\end{center}
\end{figure}

Figure \ref{fig:Fig_7_23_h} shows the corresponding plot for the convolved projected circle
LSF. In this case the steep sides and low wings of the LSF exacerbate the problem of finding
widths at pixel phase 0.5 and low sample frequencies, and as a result at 1.5 pixels/FWHM it
reaches an RMS width error almost $4 \times$ worse than the fine-sampled limit. Any sampling
frequency below 2 pixels/FWHM experiences severe noise enhancement at pixel phases close to
0.5. Even at 2.5 pixels/FWHM there is significant pixel phase $-$ dependent enhancement of the
width uncertainties.

\section{PEAK ACCURACY} 
\label{sec:peak accuracy}
This section considers the effect of pixellation on the random noise errors affecting the peak
amplitude of an unresolved spectral feature. Clarke et al. (1969) give an equation analogous to
equation (\ref{eqn:sig_x_full}) for the uncertainty of the peak amplitude, when making a
2-parameter least squares fit to an unresolved feature:

\begin{equation}
\sigma_{\rm pk}^2 = \frac{\sigma^2}{ \sum{B^2} - \Big(\sum{BB'}\Big)^2 / \sum{(B')^2} }.
\label{eqn:sig_pk_full}
\end{equation}

\begin{figure}[h]
\begin{center}
\includegraphics*[angle=0, scale=0.41]{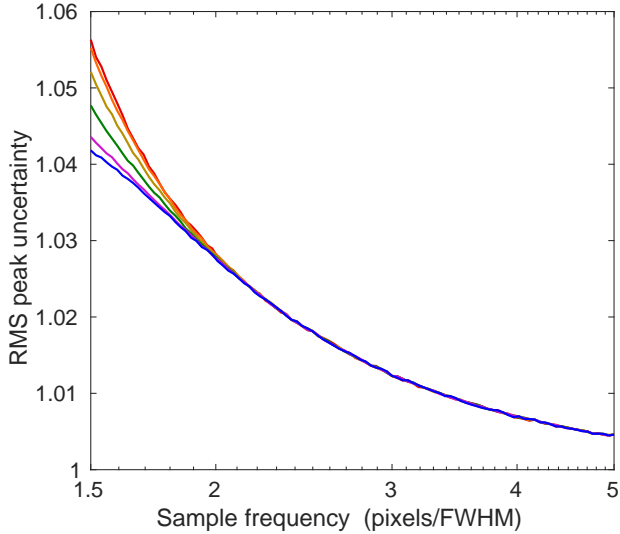} 
\caption{Peak uncertainty vs sampling frequency for the Gaussian LSF. As before, the noise is
  constant per unit wavelength interval, and is independent from one pixel to the next.  The
  uncertainties are normalised to unity at large sampling frequency, in order to show the
  effects of pixellation. The colour coding for pixel phases is as before. }
\label{fig:Fig_7_23_i}
\end{center}
\end{figure}

Figure \ref{fig:Fig_7_23_i} shows the result for a Gaussian LSF. As with wavelength (position)
and width uncertainties, pixel phase dependence develops below about 2 pixels/FWHM. However the
average effects are small, with less than 3\% increase in noise at 2 pixels/FWHM due to
pixellation. The largest errors occur for pixel phase 0.5, where the signal is spread out over
a larger effective number of pixels than for pixel phase 0.

Since $\sigma_{\rm pk}$ and the pixel noise $\sigma$ have the same dimensions, they can be
validly compared. Defining $n_{\rm eff}$ as the effective number of pixels in the `resolution
element',

\begin{equation}
\sigma_{\rm pk}^2 = \frac{\sigma^2}{n_{\rm eff}},
\label{eqn:n_eff}
\end{equation}

and equation \ref{eqn:sig_pk_full} could be used to find  $n_{\rm eff}$ .

\begin{figure}[h]
\begin{center}
\includegraphics*[angle=0, scale=0.41]{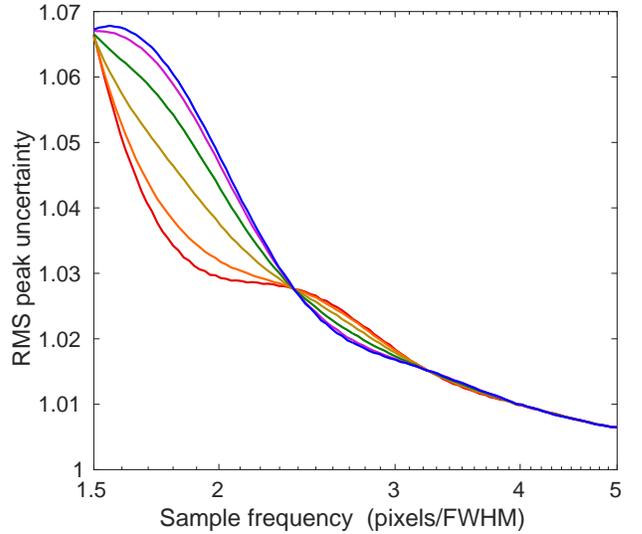} 
\caption{Peak uncertainty vs sampling frequency for the convolved projected circle LSF.
  The colour coding for pixel phases is as before. }
\label{fig:Fig_7_23_j}
\end{center}
\end{figure}

Figure \ref{fig:Fig_7_23_j} shows the corresponding change in $\sigma_{\rm pk}$ for the
convolved projected circle. Again the average increase in errors due to pixellation is small,
less than about 4\% at 2 pixels/FWHM.

\section{CONSISTENT SCALE OF SPECTRAL RESOLVING POWER}
\label{sec:consistent scale}
Sampling by finite-width pixels causes a reduction in spectral resolving power, since the
effective LSF is the unsampled LSF convolved with a rectangle equal to the pixel width. While it is
possible to calculate a resolving power loss by simply comparing the widths of the original
LSF as incident on the detector and the width after pixel convolution, this gives only
an average over pixel phases, and does not show the important differences which arise as a
function of pixel phase at low sampling frequency. Moreover, as will be shown below, it does
not give an accurate picture of the effects of pixellation on the various LSF forms.

The obvious difficulty in calculating a pixel-phase dependent resolving power lies in how
resolving power should be defined in the case of coarse sampling. The approach taken here
follows that of Paper 1, where it was argued that the modern conventional definition of
spectral resolving power as $R = \lambda / \delta \lambda$ with $\delta \lambda$ taken as the
FWHM is unsatisfactory because FWHM is a poor measure of the truly important aspects of
resolution, namely the ability to distinguish closely-spaced spectral lines, or to measure
accurate wavelengths of unresolved lines. Taking the latter property as the basis, a consistent
measure of resolution $\delta \lambda _{\sigma \lambda}= \beta$ FWHM was developed\footnote{$R
  _{\sigma \lambda} = \lambda / \delta \lambda _{\sigma \lambda}$ calculated on this scale is
  the Rayleigh criterion resolving power of an instrument with a sinc$^2$ LSF which has the
  same wavelength noise error as the instrument in question when receiving the same incident
  total line flux and subject to noise that is constant per unit wavelength interval.}. The
formula for calculation of $\beta$ was given for the case where the LSF is continuous or finely
sampled. We now consider the important case of pixellated data. (The other consistent resolving
power scale of Paper 1, based on the `$\alpha$' scaling factors, will not be considered here
because the `$\beta$' scale is easier to use in practice. However, Section \ref{sec:resolving
  features} does consider one effect of sampling on resolution of closely-spaced features.)

This approach will take into account both the broadening effect of the sampling (effectively
convolution of the LSF with the pixel rectangle) and the increase of wavelength error due to
pixellation, as seen in Figures \ref{fig:Fig_7_23_a}, \ref{fig:Fig_7_23_b},
\ref{fig:projcircleconv_sigma_lambda}, and \ref{fig:AAOmega_sigma_lambda}. In this context
`wavelength uncertainty' refers to random noise errors - the bias errors which also depend on
pixel phase but which are not diminished by high signal/noise data are considered later.

The analysis requires knowledge of the `intrinsic LSF', which is used to mean the continuous
LSF that has fallen on the detector, before sampling. This is known in the case of model LSFs
such as the Gaussian, Lorentzian and the convolved projected circle, and could also apply to
smooth model LSFs derived from optical design software models or an empirical LSF processed as
for the AAOmega LSF above (with deconvolution to remove the effects of pixel sampling, at least
approximately).  With the intrinsic LSF known, the effects of any proposed sampling frequency
can be evaluated.

Following Paper 1 the `$\beta$' scale of consistently-defined resolving power is based
on equating the wavelength errors for the LSF in question and that of a sinc$^2$ LSF, {\it
  i.e.}

\begin{equation}
\sigma_{{\lambda}\ ({\rm sinc}^2,{\rm cont})}  = \sigma_{{\lambda}\ ({\rm LSF,pix,phase})}
\label{eqn:sig_lam_equality}
\end{equation}

where the left-hand side represents the $\sigma_{\lambda}$ of a sinc$^2$ LSF that is continuous
or very finely sampled, while the right hand side is the $\sigma_{\lambda}$ of the LSF under
study, which is pixellated and observed at some particular pixel phase. Thus given a certain
LSF form, sampling frequency and pixel phase, equation \ref{eqn:sig_lam_equality} can be used
to give the FWHM of the sinc$^2$ profile required to satisfy the equality.

The analysis will closely follow that of Paper 1. Since a variety of pixel widths are now
considered it is necessary to write the spectral noise as 

\begin{equation}
\sigma_1  = \sigma \sqrt{\Delta \lambda}.
\label{eqn:sig_1}
\end{equation}

Here $\sigma_1$ is the noise for unit wavelength (dispersion axis) interval and is constant
both within a spectrum and between the two LSFs considered, while $\sigma$ is then the noise in
a pixel (wavelength channel) of width $ \Delta \lambda$, and is assumed to be uncorrelated
between pixels.

In order to use the Clarke et al. formulas (equations \ref{eqn:sig_x_full} or
\ref{eqn:sig_x_simple}) it is convenient to define $S$ by

\begin{equation}
1/S =\Delta \lambda  \Big( \sum{(B')^2} - \Big(\sum{BB'}\Big)^2 /  \sum{B^2} \Big)         
\label{eqn:S_full}
\end{equation}

or, if the simple formula is adequate,

\begin{equation}
1/S =\Delta \lambda  \Big( \sum{(B')^2}  \Big).         
\label{eqn:S_simple}
\end{equation}

$S$ will play the role for pixellated data that the  `noise width',

\begin{equation}
Z = \frac{1}{\int_{-\infty}^{+\infty}(B')^2 d \lambda}.
\label{eqn:Z}
\end{equation}

plays for continuous symmetric LSFs. The crucial difference is that $S$ depends on the sample frequency
and pixel phase as well as on the LSF form. Following the same procedure as in Paper 1, {\it
  i.e.} equating the RMS wavelength errors for the pixellated LSF and a continuous sinc$^2$ LSF
under the condition of equal fluxes (peak $\times$ equivalent width) the result is 

\begin{equation}
\beta = 1.3809 \ S^{\frac{1}{3}}_{\rm LSF} {\rm EW}^{\frac{2}{3}}_{\rm LSF}   /  {\rm FWHM}_{\rm LSF}     
\label{eqn:beta_1}
\end{equation}

where EW stands for the equivalent width (area/peak height).
Hence the resolution element which should be used in place of FWHM$_{\rm LSF}$ is 

\begin{equation}
\delta \lambda_{\sigma \lambda} = \beta \ {\rm FWHM}_{\rm LSF}  = 1.3809 \ S^{\frac{1}{3}}_{\rm LSF} {\rm
  EW}^{\frac{2}{3}}_{\rm LSF}  .
\label{eqn:resoln_elt}
\end{equation}

The value of $\beta$ reflects the effects of pixellation, LSF
form and conversion to the Rayleigh criterion of a sinc$^2$ profile to define what `just
resolved' means. To calculate the final resolving power on the consistently-defined scale, use

\begin{equation}
R_{\sigma \lambda} = \frac{1}{\beta} \frac{\lambda}{{\rm FWHM}_{\rm LSF}}.
\label{eqn:R}
\end{equation}

\begin{figure}[h]
\begin{center}
\includegraphics*[angle=0, scale=0.41]{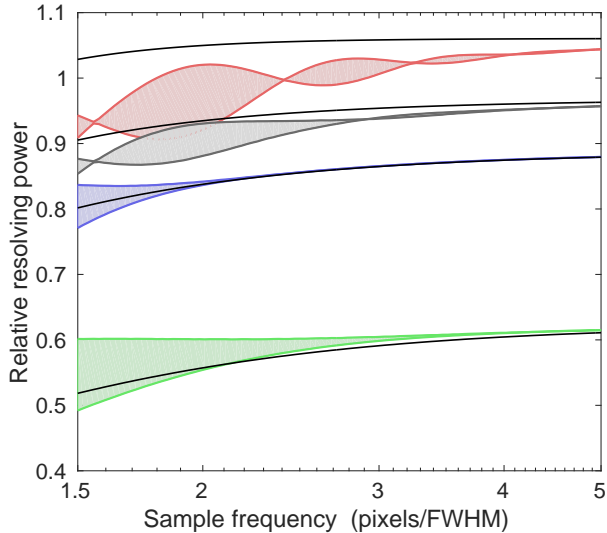}
\caption{Relative resolving power $R_{\sigma \lambda}/R = 1/\beta$, where $R_{\sigma \lambda}$
  is the resolving power of the LSF in question, as subject to pixellation and measured on the
  consistently defined `$\beta$' scale, and $R = \lambda/{\rm FWHM}$ is the
  conventionally-defined resolving power of the LSF, in the limit of fine sampling. The plot
  shows the relative values of $R_{\sigma \lambda}/R$ for the four LSFs, when all have the same
  unsampled FWHM.  On this scale, the value 0.886 corresponds to the resolving power of a
  fine-sampled sinc$^2$ LSF (again with the same FWHM), using the Rayleigh criterion to define
  resolution.  The shaded areas indicate the range covered by different pixel phases. From the
  top the curves are: red - convolved projected circle; grey - AAOmega LSF; blue - Gaussian;
  green - Lorentzian. The black curves show the effect of pixel convolution on the conventional
  $R = \lambda/{\rm FWHM}$.}
\label{fig:Fig_7_23_k}
\end{center}
\end{figure}

Figure \ref{fig:Fig_7_23_k} shows the results for the four LSFs discussed in section
\ref{sec:wavelength accuracy}. Rather than assume some arbitrary resolving power for
comparison, they are presented as relative resolving power, which is just $1/\beta$ as
calculated from equation \ref{eqn:beta_1}. The factors which affect this plot are: 1) For a
given FWHM, the basic LSF form affects the resolving power when measured on the consistent
scale which is based on equality of $\sigma_{\lambda}$. This is the reason the Lorentzian's
values are low, while the convolved projected circle is high; 2) Sampling by finite width
pixels reduces the resolving power by increasing the wavelength uncertainties as the sample
frequency is reduced; 3) Pixel phase becomes increasingly important at low sample frequencies;
4) The absolute values on the vertical scale are relative to the value 0.886 (=1/1.129) which
applies to a sinc$^2$ LSF with fine sampling and using the Rayleigh criterion to define `just
resolved'. 

The black curves show the result of calculating the resolving power reduction due to sampling
by simply convolving the LSF with the pixel rectangle and comparing the FWHM with that of the
intrinsic LSF. For convenience, the values have been scaled to agree with the corresponding
consistently-defined relative resolving power at the limit of fine sampling. Only for the
Gaussian LSF does this simplistic calculation show approximately the correct dependence on
sample frequency. For the others it is a poor approximation, especially for the convolved
projected circle or the AAOmega LSF - which represent forms encountered in any spectrograph fed
by multi-mode fibres. Furthermore, the simplistic calculation is unable to take account of the
dependence on pixel phase.

%\sloppy
As an example, $R_{\sigma \lambda}$ has been calculated for the AAOmega configuration above,
using the deconvolution of the empirical fit to the LSF (shown in Figure \ref{fig:Fig_7_23_e}) and
the known centre wavelength (725.21 nm) and dispersion (0.1568 nm/pixel). The above process
gives $R_{\sigma \lambda} = 1282.8 - 1286.9$, with the range of values being due to different
pixel phases. The range is narrow because the profile is well sampled. For comparison, the
conventional $R = \lambda / {\rm FWHM_{\rm LSF}} = 1358.7$. In this case the difference between the
conventional $R$ and $R_{\sigma \lambda}$ is small ({\it i.e} $\beta$ is close to 1), due to
two competing effects: the LSF has steeper sides than a sinc$^2$ profile, which will raise
$R_{\sigma \lambda}$, but then the result is scaled down to give resolving power equivalent
to that from the Rayleigh criterion. There is little loss of resolving power due to pixellation
because the sampling frequency is 3.41 pixels/FWHM.
%\fussy

\section{SYSTEMATIC BIAS ERRORS} 
\label{sec:bias}
As well as the increased random errors described above in Sections \ref{sec:wavelength
  accuracy}, \ref{sec:width accuracy} and \ref{sec:peak accuracy}, pixellation of a spectrum
can also lead to bias errors. In general such errors will depend on the pixel phase of a
spectral feature. They are insidious because they differ from the usual noise errors in that
they are not reduced by high signal/noise ratio, and thus must eventually dominate (perhaps
with other systematic errors) for very high S/N spectra. In that case they would set a
quasi-random noise floor, when considering an ensemble of spectral features of various pixel
phases.  Such errors will occur if the spectral features are fitted using a functional form
which does not exactly match the LSF, such as the common practice of fitting (say) a Gaussian
to features which are not exactly Gaussian. There are also small bias errors in using the
model-independent centroid as a location parameter.

Bias errors in the flux of a spectral line can in principle be easily avoided simply by summing
the contributions of the wavelength channels (pixels) which include the line, but this is a
noisy process, subject to truncation error in the line wings. Flux bias will be introduced if
an incorrect functional form is fitted.

%Errors in position (wavelength), however, are a significant issue. Errors
%are in general small, as will be shown below, but for high-resolution stable spectrographs now
%carrying out planetary search observations, even small systematic errors must be considered. 

\subsection{Exact LSF fitted to spectral features}
\label{sec:Exact LSF}
It is not possible to give a general formula which includes all possible LSFs and fitting
functional forms. In general the only approach is to carry out simulations, and a number of
these will be presented below. The first case to consider is when the exactly correct
functional form is fitted, {\it i.e.} the intrinsic LSF is known exactly, and it is integrated
over samples for every spectral feature, taking due account of the pixel phase. No plots will
be shown for this case because it results in bias-free fitting, even when there is a degree of
undersampling ({\it i.e.} the LSF contains Fourier components beyond the Nyquist limit).
% should give evidence/example?

\subsection{Gaussian fits}
\label{sec:gaussian fits}
A common practice in spectral analysis is to fit a Gaussian profile to spectral lines, on the
basis that although the LSF is not exactly Gaussian, it is likely to be fairly close. This
section examines the resulting biases.

\begin{figure}[h]
\begin{center}
\includegraphics[angle=0, scale=0.40]{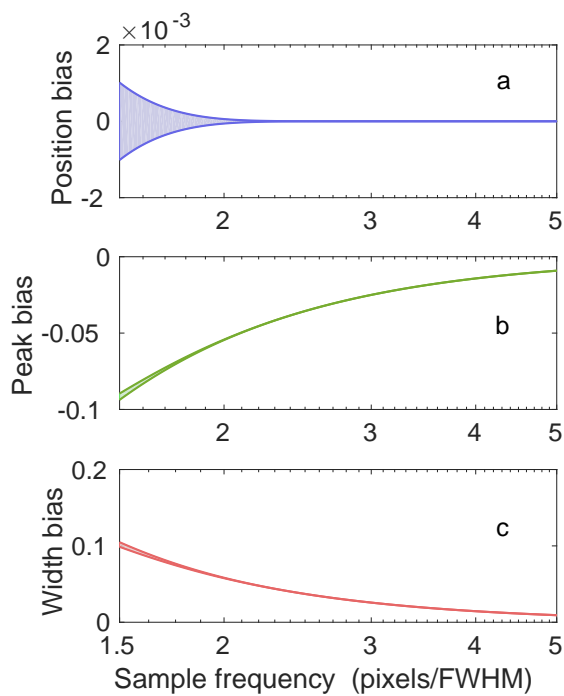}
\caption{Bias errors of position, peak height and width for a plain Gaussian fitted to a
  sampled Gaussian LSF. For position and width the errors are relative to the FWHM of 1.0, and
  the peak bias is relative to the Gaussian LSF peak = 1.0. The filled areas show the range of
  values covered by different pixel phases. The annotation `$\times 10^{-3}$' on the vertical
  axis of panel `a' applies only to that panel. }
\label{fig:Fig_6_20c}
\end{center}
\end{figure}

The first case is that of a Gaussian intrinsic LSF, but the fitting function is a plain
point-sampled Gaussian, {\it i.e.} not integrated across samples. Thus there are misfit errors
which are worse for lower sampling frequency. Note that it is necessary to make a 3-parameter
fit, allowing a variable width as well as position and peak height, because the fitted form is
not the correct one. Figure \ref{fig:Fig_6_20c} shows the results for each of the three
parameters. The position bias increases rapidly below 2 pixels/FWHM, reaching pixel-phase
dependent extremes of $\pm 0.0010$ (0.1\% of the FWHM) at 1.5 pixels/FWHM. Such errors would be
unimportant in low to moderate signal/noise data, and the crude fit of an inappropriate LSF
would not be used in high precision work.  The bias errors in peak height and width are
substantial at low sampling frequencies, with the effect of functional form mismatch
overshadowing the effect of pixel phase. With the peak underestimated and the width
overestimated by a similar factor, the flux error is much reduced and reaches only 0.13\% at
1.5 pixels/FWHM.

\begin{figure}[h]
\begin{center}
\includegraphics[angle=0, scale=0.40]{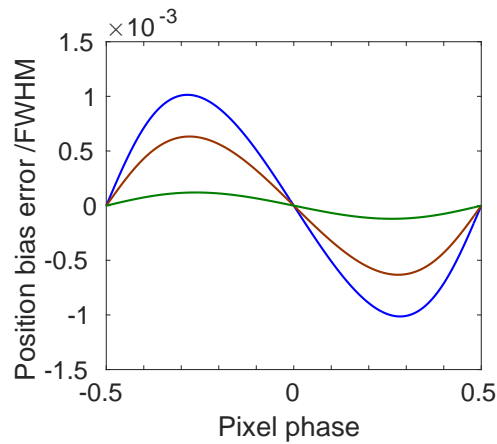}
\caption{Bias errors of position as a function of pixel phase, for a plain Gaussian fitted to a
  Gaussian LSF. The highest amplitude curve is for sample frequency of 1.5 pixels/FWHM, the
  others are at 1.6 and 1.9 pixels/FWHM.}
\label{fig:Fig_6_21}
\end{center}
\end{figure}

The data of Figure \ref{fig:Fig_6_20c}a are presented as a function of pixel phase in Figure
\ref{fig:Fig_6_21}. As expected, the position error is zero when the pixels are symmetric with
respect to the peak, at phases 0 and $\pm 0.5$, and varies approximately sinusoidally at other
pixel phases. The error decreases rapidly with increasing sample frequency, and is negligible
above 2 pixels/FWHM.

\begin{figure}[h]
\begin{center}
\includegraphics[angle=0, scale=0.40]{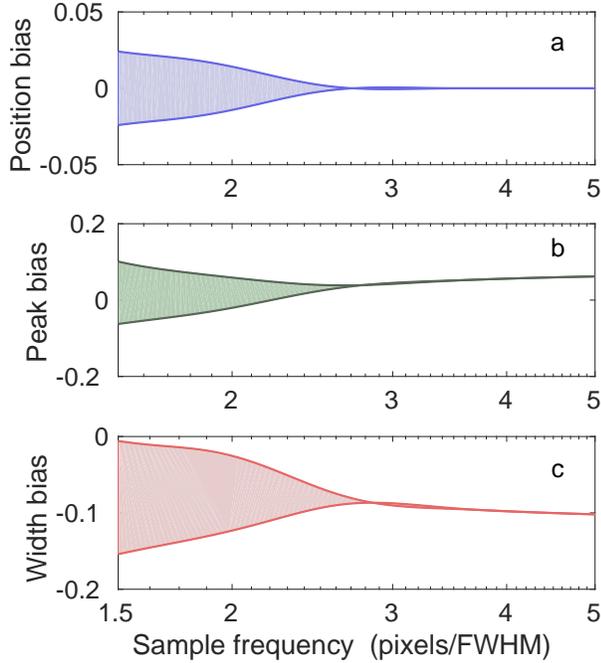}
\caption{Bias errors of position, peak height and width for a plain Gaussian fitted to an LSF
  derived from convolution of a projected circle with a Gaussian that gives the minimum final
  intrinsic FWHM. For position and width the errors are relative to the FWHM of 1.0, and the
  peak bias is relative to the intrinsic LSF peak = 1.0. The filled areas show the range of
  values covered by different pixel phases. }
\label{fig:Fig_6_22}
\end{center}
\end{figure}

The next LSF considered is the projected circle convolved with the Gaussian which results in
the minimum final intrinsic FWHM, as used in Figure \ref{fig:Fig_6_54_prjcrcconv_b}. The results
for biases in position, peak and width for this case are shown in Figure \ref{fig:Fig_6_22}.
The position bias errors are much larger than for the Gaussian LSF (at sample frequency 1.5 the
range is $24 \times$ greater), and extend to beyond 2 pixels/FWHM. The peak and width errors
also have greatly increased pixel-phase dependence, and unlike Figure \ref{fig:Fig_6_20c} the
biases do not asymptote to zero at high sample frequency because in this case the Gaussian fit
is inherently an incorrect functional form. Quite apart from issues of pixellation, this is an
illustration of the danger of using an inappropriate functional form, since these errors would
be apparent in even moderate S/N data. For calculation of the flux (area) the peak and width
errors again tend to compensate, and the maximum flux error is +3.5\% at 2.21 pixels/FWHM.

\begin{figure}[h]
\begin{center}
\includegraphics[angle=0, scale=0.40]{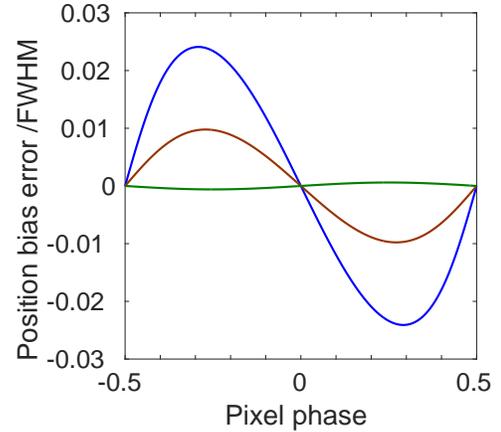}
\caption{Bias errors of position as a function of pixel phase, for a plain Gaussian fitted to
  the convolved projected circle LSF. The highest amplitude curve is for sample frequency of
  1.5 pixels/FWHM, the others are at 2.16 and 2.96 pixels/FWHM. }
\label{fig:Fig_6_23}
\end{center}
\end{figure}

\begin{figure}[h]
\begin{center}
\includegraphics[angle=0, scale=0.40]{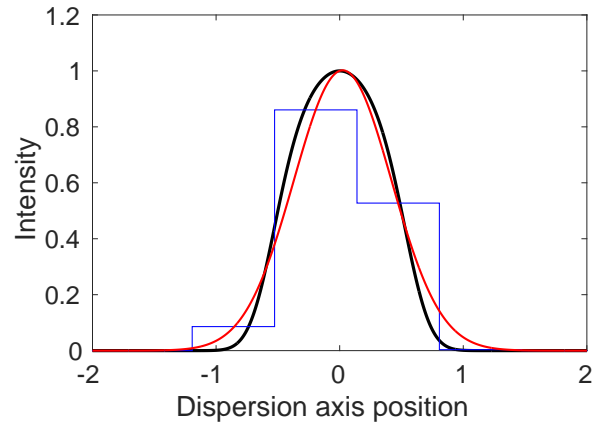}
\caption{Example of the misfit of a pure Gaussian to the convolved projected circle. Thick
  black line: intrinsic convolved projected circle LSF; blue `histogram' plot: the LSF sampled at 1.5
  pixels/FWHM and pixel phase -0.29, which gives the maximum position bias error; red curve:
  the Gaussian which best fits the sampled data.}
\label{fig:Fig_6_23b}
\end{center}
\end{figure}

Figure \ref{fig:Fig_6_23} shows the position bias errors vs pixel phase. Again an approximately
sinusoidal form is seen, but with much larger amplitude, which could affect moderate S/N
spectra. Figure \ref{fig:Fig_6_23b} shows one example of the misfit which occurs when
attempting to fit this LSF with a plain Gaussian.

\begin{figure}[h]
\begin{center}
\includegraphics[angle=0, scale=0.40]{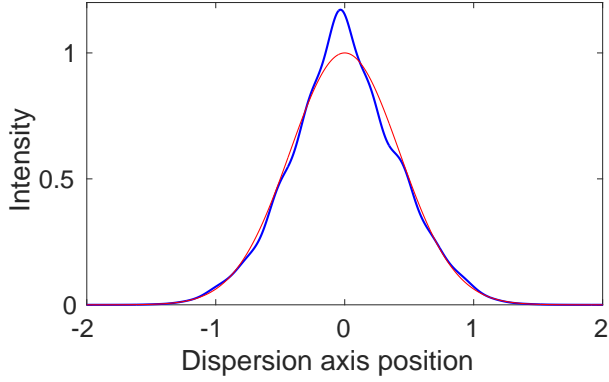}
\caption{Thick blue curve - Gaussian LSF perturbed by some high-frequency noise. Thin red curve
  - the unperturbed parent Gaussian LSF. The perturbation is a sum of 3 sine curves with
  amplitudes $A_i = [0.1,0.05,0.03]$, pixel phases at sine wave zero crossing $\phi_i =
  [-0.33,0.85,-0.6]$ and frequencies $f_i = [1,2,4]$ where $f_1$ corresponds to one cycle
  across the FWHM.  The final perturbed curve is a Gaussian of peak and FWHM = 1 multiplied by
  (1 + the sum of sine waves).}
\label{fig:Fig_6_27_1}
\end{center}
\end{figure}

\begin{figure}[h]
\begin{center}
\includegraphics[angle=0, scale=0.40]{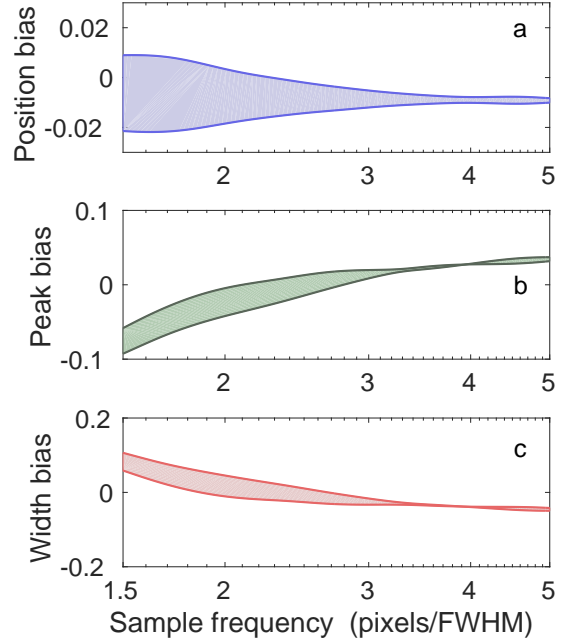}
\caption{Bias errors of position, peak height and width for a plain Gaussian fitted to a
  perturbed Gaussian LSF. For position and width the errors are relative to the parent FWHM of
  1.0, and the peak bias is relative to the parent Gaussian LSF peak = 1.0. The filled areas
  show the range of values covered by different pixel phases. The errors have been calculated
  relative to the original unperturbed LSF, and therefore have essentially arbitrary zero point
  offsets.}
\label{fig:Fig_6_26_pert}
\end{center}
\end{figure}

\begin{figure}[h]
\begin{center}
\includegraphics[angle=0, scale=0.40]{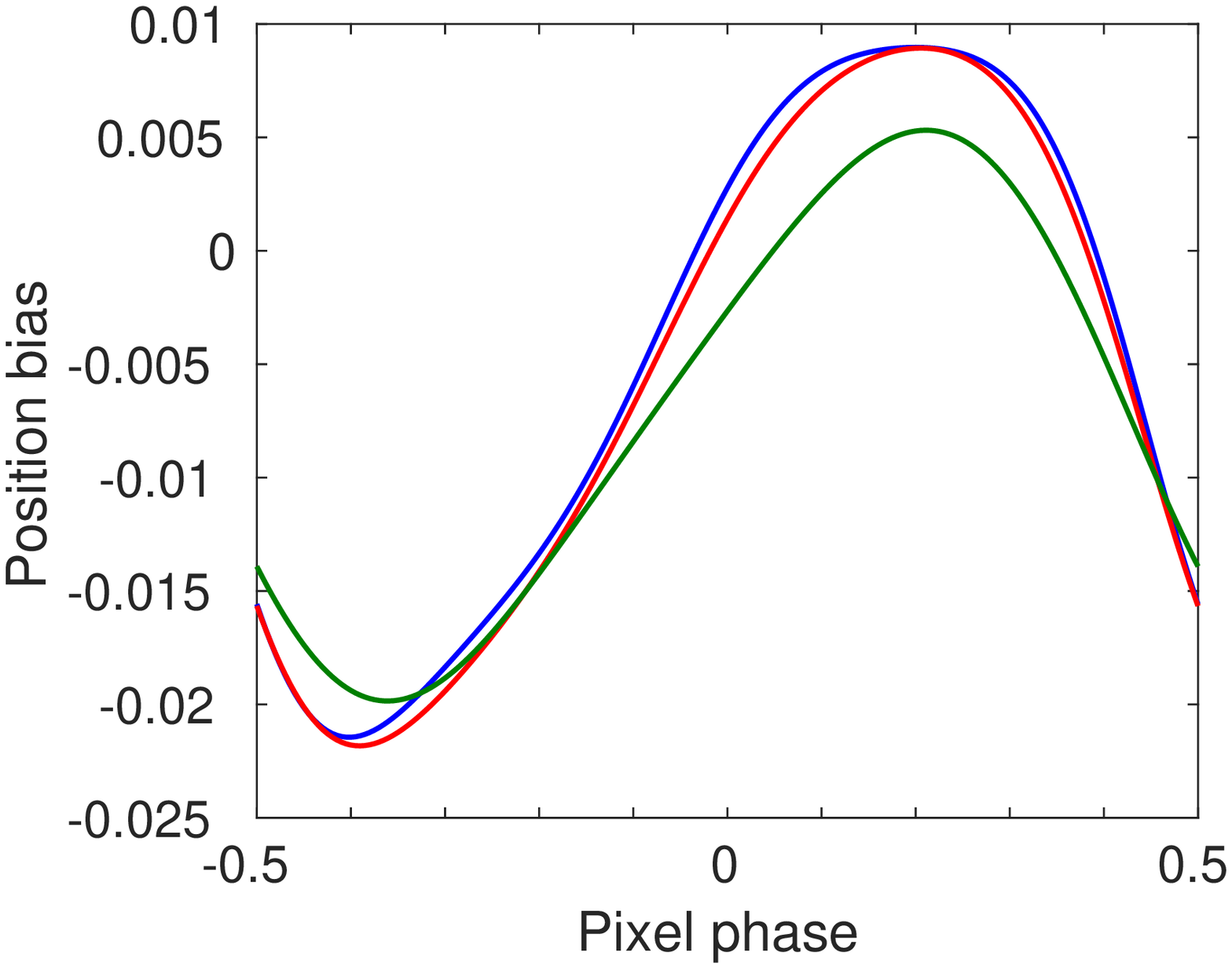}
\caption{Bias errors of position as a function of pixel phase, for a plain Gaussian fitted to
  the perturbed Gaussian LSF. The blue curve is for sample frequency of 1.5
  pixels/FWHM, red is at 1.6 and green at 1.9 pixels/FWHM.}
\label{fig:Fig_6_27_2}
\end{center}
\end{figure}

The above two LSFs are perfectly symmetrical, an ideal which is not achieved in practice, due
to complex residual optical aberrations and other effects. These can give rise to fine
structure in the LSF, which is of particular importance because LSF structure finer
than the pixel scale will cause a shift in the fitted location of the feature, as it moves from
contributing in one pixel to the next one. It is an example of the desirability of band
limiting the data {\it before} sampling, which is not possible in this case. Thus
the smoothing effect of integrating over pixels does not reduce this bias.  The final LSF to
be considered in this section is an example of the difference that some asymmetric fine
structure can make to the bias errors. Figure \ref{fig:Fig_6_27_1} shows an LSF constructed by
perturbing a Gaussian with three sinusoids, having spatial periods of 1, 0.5 and 0.25 $\times$
the FWHM. Figures \ref{fig:Fig_6_26_pert} and \ref{fig:Fig_6_27_2} show the results. Compared
with the fits to the pure Gaussian LSF in Figure \ref{fig:Fig_6_20c} the bias errors in
position are much larger (15$\times$ greater at sample frequency 1.5) and continue to much
larger sampling frequencies, despite the perturbation of the Gaussian form being relatively minor. The
bias errors in these Figures were computed relative to the unperturbed parent Gaussian, thus
the zero points on the vertical scales are essentially arbitrary, and attention should be
focussed on the range and variation of the errors. The flux (area) errors are not large, with
the greatest being -1.8\% at 2.1 pixels/FWHM.

In Figure \ref{fig:Fig_6_27_2} the same three sample frequencies are plotted as for the pure
Gaussian case (Figure \ref{fig:Fig_6_21}); it is notable that the position errors are much
larger and do not decrease nearly as quickly as sample frequency increases. Thus the smoothness
of the {\it intrinsic} (unsampled) LSF is an important criterion if accurate wavelengths are to
be obtained from a spectrograph when using simple line fitting procedures.
% ref MJI GHOST?

\subsection{Centroid position}
\label{sec:centroid position}
The centroid (centre of gravity) of an isolated spectral feature is 
\begin{equation}
C_T = \frac{\int_{-\infty}^{\infty}F(x)x\ dx}{\int_{-\infty}^{\infty}F(x)dx}
\label{eqn:centroid_T}
\end{equation}
where $F(x)$ is the spectral intensity as a function of location $x$ along the dispersion axis,
and $C_T$ represents the true centroid, {\it i.e} free from the effects of pixellation. This
formula assumes that any background level has been subtracted, so the peak is sitting on zero
background level. For pixellated data one has to use
\begin{equation}
C_P = \frac{\sum F_i x_i}{\sum F_i}
\label{eqn:centroid_P}
\end{equation}
where the sum is over all pixels containing any part of the peak. The difference of these two
expressions represents the bias of the pixellated centroid:
\begin{equation}
C_P - C_T = \frac{\int_{-\infty}^{\infty}F(x)(x_i - x)\ dx}{\int_{-\infty}^{\infty}F(x)dx}
\label{eqn:centroid_bias}
\end{equation}
where $x_i$ is the centroid of pixel $i$ which is nearest to the particular value of $x$. For
pixels assumed to have uniform sensitivity, $x_i$ is the $x$ coordinate value at the centre of
the pixel. The $(x_i - x)$ factor in this equation imparts a sawtooth characteristic to the
integrand, which results in near-cancellation of the positive and negative parts. The overall
result is further reduced by near cancellation of the contributions from the ascending and
descending slopes of the LSF. The result is that centroid bias values can be very small under
some circumstances. The bias results for some example LSFs will be considered below. 

The principal advantage of the centroid is that it is model-independent, providing that it is
clear where in the wings of the feature the summation should be truncated. The outstanding
disadvantage is the centroid's noise response. In the case of uniform noise of standard
deviation $\sigma$ in each pixel, it can be shown that the variance of the pixellated centroid
is given by
\begin{equation}
\sigma _{C_P}^2 = \frac{\sigma^2}{(\sum F_i)^2}  \sum (x_i - C_P)^2,
\label{eqn:centroid_noise}
\end{equation}
which results in the variance increasing quadratically and without limit as more pixels are
included in the wings of the feature. In this respect least squares fits to the LSF are
obviously superior. However, the centroid is still of interest for the reasons given above, and
in fact if the summation for centroid evaluation is suitably truncated, its noise need not be
excessive: for example in the case of a Gaussian LSF the centroid noise given by equation
\ref{eqn:centroid_noise} equals the 2-parameter least squares fit noise (equations
\ref{eqn:sig_x_full} or \ref{eqn:sig_x_simple}) when the centroid summation includes 98 - 99 \%
of the flux.

\begin{figure}[h]
\begin{center}
\includegraphics[angle=0, scale=0.395]{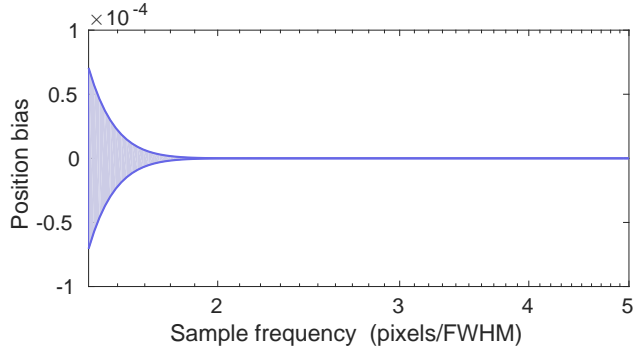}
\caption{Bias error of centroid positions for a Gaussian LSF. The filled area shows the range
  covered by different pixel phases.}
\label{fig:Fig_6_35}
\end{center}
\end{figure}
\begin{figure}[h]
\begin{center}
\includegraphics[angle=0, scale=0.3]{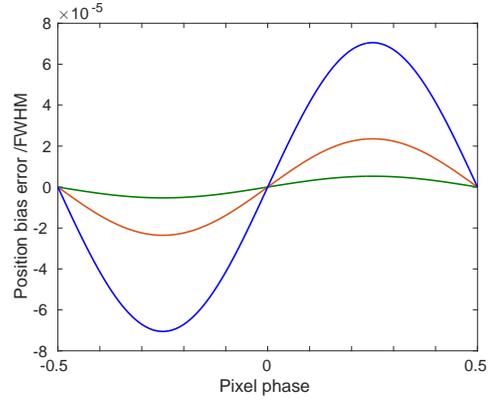}
\caption{Bias errors of centroid position as a function of pixel phase for a Gaussian LSF. The
  largest amplitude curve (blue) is at 1.50 pixels/FWHM; the other two are at 1.594 and 1.715
  pixels/FWHM.}
\label{fig:Fig_6_35b}
\end{center}
\end{figure}

Figure \ref{fig:Fig_6_35} shows the centroid bias for the case of a Gaussian LSF. Bias is
negligible at 2.0 pixels/FWHM, and even at 1.5 pixels/FWHM it reaches a maximum (at particular
pixel phases) of only $7.05 \times 10^{-5}$ of the FWHM. The behaviour as a function of pixel
phase is shown in Figure \ref{fig:Fig_6_35b}, illustrating the near-sinusoidal form of the very
small bias errors. The maximum bias is $14 \times$ less than that shown in Figure
\ref{fig:Fig_6_20c} for the fit of a plain Gaussian to a Gaussian LSF, illustrating the
bias-resistant nature of the centroid in this case.

\begin{figure}[h]
\begin{center}
\includegraphics[angle=0, scale=0.41]{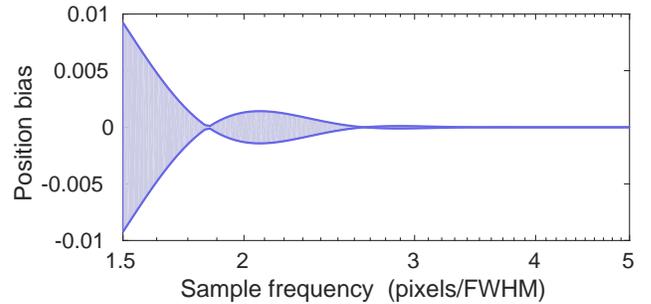}
\caption{Bias error of centroid positions for the convolved projected circle LSF. The filled
  area shows the range covered by different pixel phases.}
\label{fig:Fig_6_37b}
\end{center}
\end{figure}

\begin{figure}[h]
\begin{center}
\includegraphics[angle=0, scale=0.40]{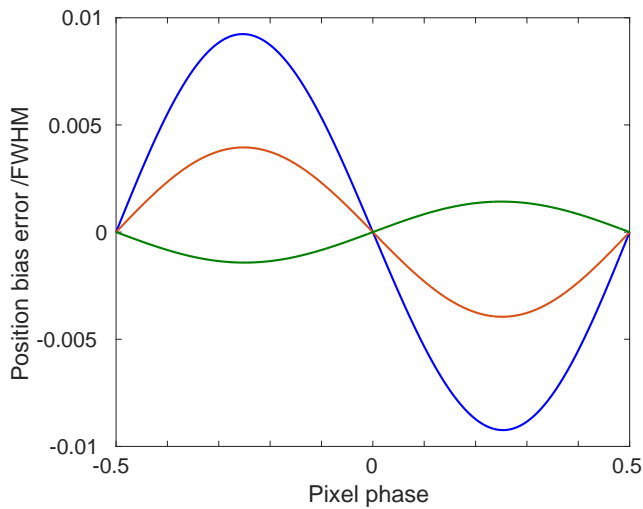}
\caption{Bias errors of centroid position as a function of pixel phase for the convolved
  projected circle LSF.  The largest amplitude curve (blue) is at 1.50 pixels/FWHM; the other
  two are at 1.653 and 2.083 pixels/FWHM.}
\label{fig:Fig_6_37}
\end{center}
\end{figure}

For the convolved projected circle LSF the steeper sides and flattened top produce much larger
errors, as shown in Figures \ref{fig:Fig_6_37b} and \ref{fig:Fig_6_37}. At 1.5 pixels/FWHM the
maximum error is $9.24 \times 10^{-3}$, {\it i.e.} $131 \times$ greater than for the Gaussian
LSF. Furthermore, there is a secondary maximum in the bias errors at 2.08 pixels/FWHM, where
the largest error is $1.42 \times 10^{-3}$.

\begin{figure}[h]
\begin{center}
\includegraphics[angle=0, scale=0.40]{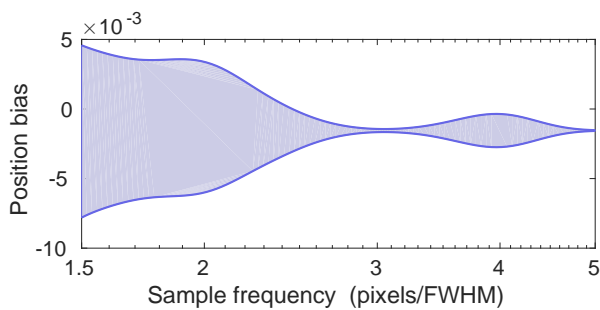}
\caption{Bias error of centroid positions for the perturbed Gaussian LSF. The filled area shows the range
  covered by different pixel phases.}
\label{fig:Fig_6_38}
\end{center}
\end{figure}

The final LSF form considered here is the perturbed Gaussian, as shown in Figure
\ref{fig:Fig_6_27_1}.
% used above in Figures
% \ref{fig:Fig_6_26_pert} and \ref{fig:Fig_6_27_2}. 
The result for its centroid bias as a function of sample frequency is shown in Figure
\ref{fig:Fig_6_38}. The errors are far larger than for the centroid of a pure Gaussian - with
the maximum range of errors at 1.5 pixels/FWHM being $1.24 \times 10^{-2}$, which is $88
\times$ greater than the range for the pure Gaussian at the same sample frequency. This shows
the great importance of high frequency distortion of the LSF for centroid bias. The presence of
high frequency components in the LSF also produces (in this example) a secondary maximum in
bias range at a sampling frequency as high as 4 pixels/FWHM, where the range is $2.4 \times
10^{-3}$. Other tests (not shown) used an asymmetrical Gaussian, constructed from two
half-Gaussians of equal peak height but unequal widths, joined at the peak: for FWHMs differing
by a factor of 1.10 the centroid bias was $22 \times$ greater than for the symmetrical Gaussian
and for a width ratio of 1.44 the bias was $160\times$ greater (at 2 pixels/FWHM).

\section{RESOLVING CLOSELY-SPACED FEATURES} 
\label{sec:resolving features}
The consistent scale of resolving power described in  section \ref{sec:consistent scale}
provides a quantitative measure which takes into account the effect of finite pixel
widths. Nevertheless, users of pixellated spectra are likely to want further information
regarding how much the pixellation affects the ability to distinguish two closely-spaced
spectral lines. There are many ways one might quantify the effects of pixellation - for example
the increase in flux or wavelength uncertainty of one spectral line as another draws
closer. However, there are a large number of variable parameters and it is not clear that such
results would be helpful. Perhaps the most basic property of resolved lines that an observer
looks for is the presence of a relative minimum between the two peaks (in the case of emission
lines). In this section, the effects of pixellation on such a relative minimum are examined.

\begin{figure*}
\begin{center}
\includegraphics[angle=0, scale=0.40]{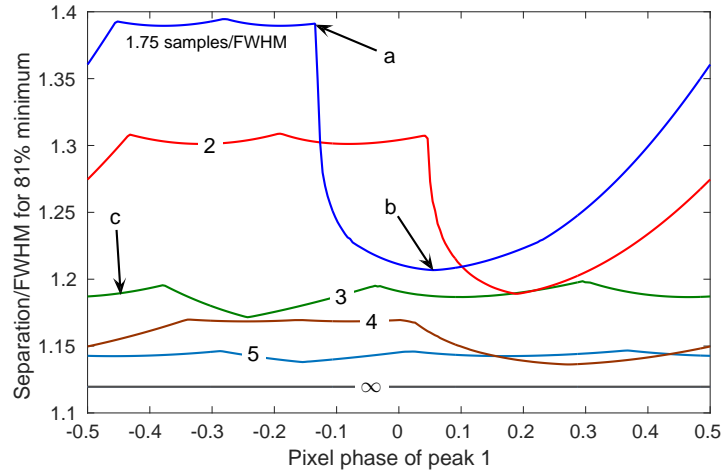}
\caption{Illustration of the effects of pixellation on the ability to see two equal height
  Gaussian peaks as separate. The horizontal axis gives the pixel
  phase of the first of the two peaks. The  vertical axis shows the
  separation of the two peaks that is required for there to be a local minimum in the
  pixellated data which is 81.1\% of the lower of the two pixellated main peaks. Curves are
  given for sampling frequency values  from 1.75 to 5 pixels/FWHM, as
  labelled. The  horizontal grey line at separation/FWHM
  = 1.1196 is the limiting separation in the case of finely sampled LSFs. The points labelled
  `a',`b',`c' refer to sample frequencies and pixel phases whose LSFs are shown in Figure
  \ref{fig:Fig_6_65a}.}
\label{fig:Fig_6_164}
\end{center}
\end{figure*}

Following the standard adopted in Paper 1, the criterion for two spectral features that are
individually unresolved to be regarded as just resolved from each other is that there should be
a relative minimum between the two (intrinsically equal) peaks that is 81.1\% of the height of either
peak. This is based on the Rayleigh criterion separation of two sinc$^2$ LSFs and has the
advantage that the criterion itself is independent of the noise properties of the data.  The
procedure adopted was to first choose an LSF functional form and sampling frequency, and then
place the first peak at the origin with a specified pixel phase. An iterative
procedure was then used to find the separation of the two (intrinsic) peaks which results in
the relative minimum of the sampled data being 81.1\% of the height of the lower of the two
sampled main peaks (they will in general be unequal due to the sampling). Figure
\ref{fig:Fig_6_164} shows the results for Gaussian LSFs. The curves have a complex structure
because changing the separation changes the pixel phase of the second peak, which in turn
affects the desired separation if the pixellated second peak is the lower of the two. Changing
separation also changes the contribution to the first peak from the wings of the second
peak. The curves are not symmetric about zero pixel phase of the first peak because the second
peak lies towards positive $x$ values and so contributes predominantly on that side. The
 horizontal grey line indicates the separation in the limit of fine sampling.

\begin{figure*}
\begin{center}
\includegraphics[angle=0, scale=0.22]{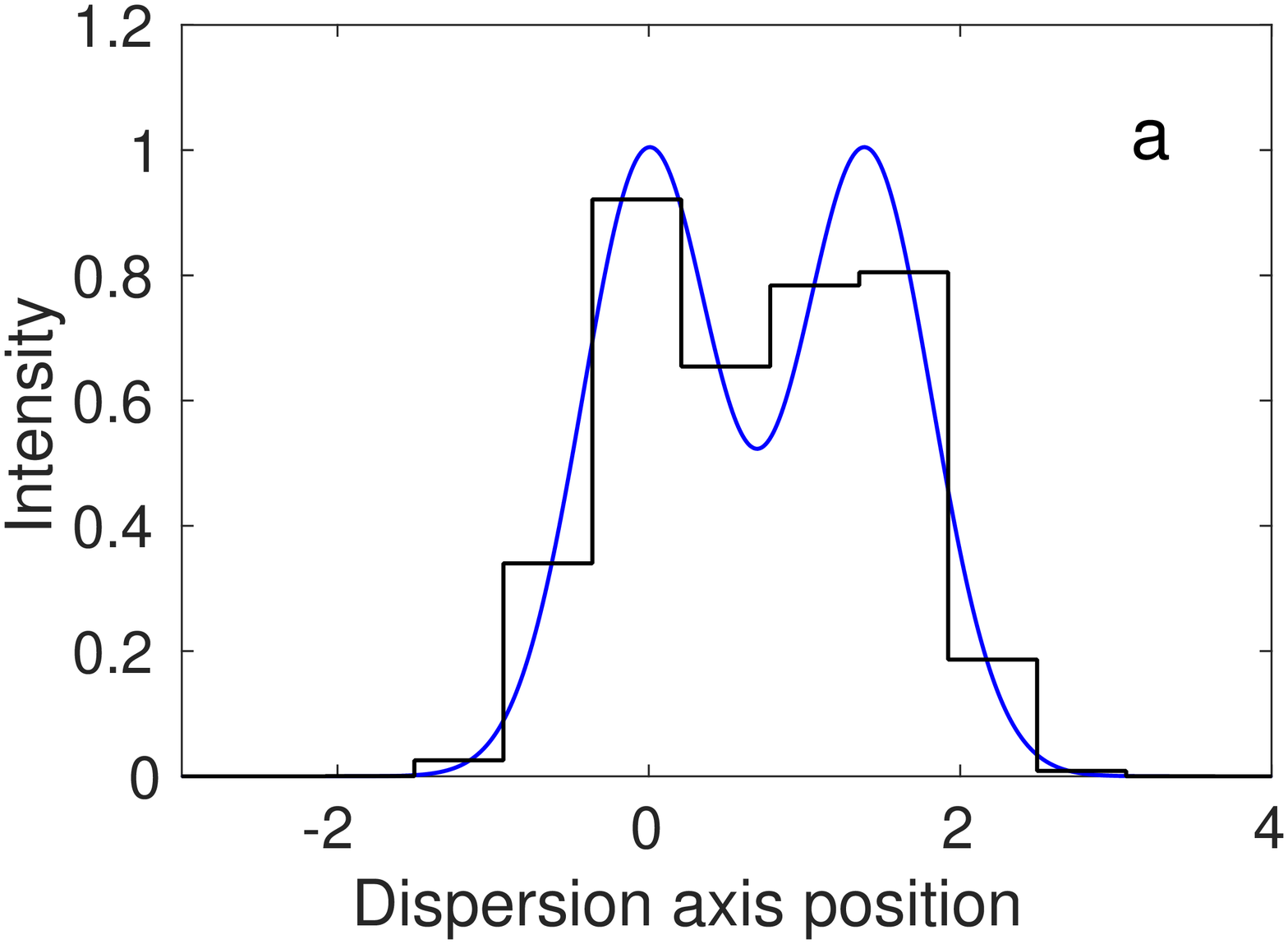}
\includegraphics[angle=0, scale=0.22]{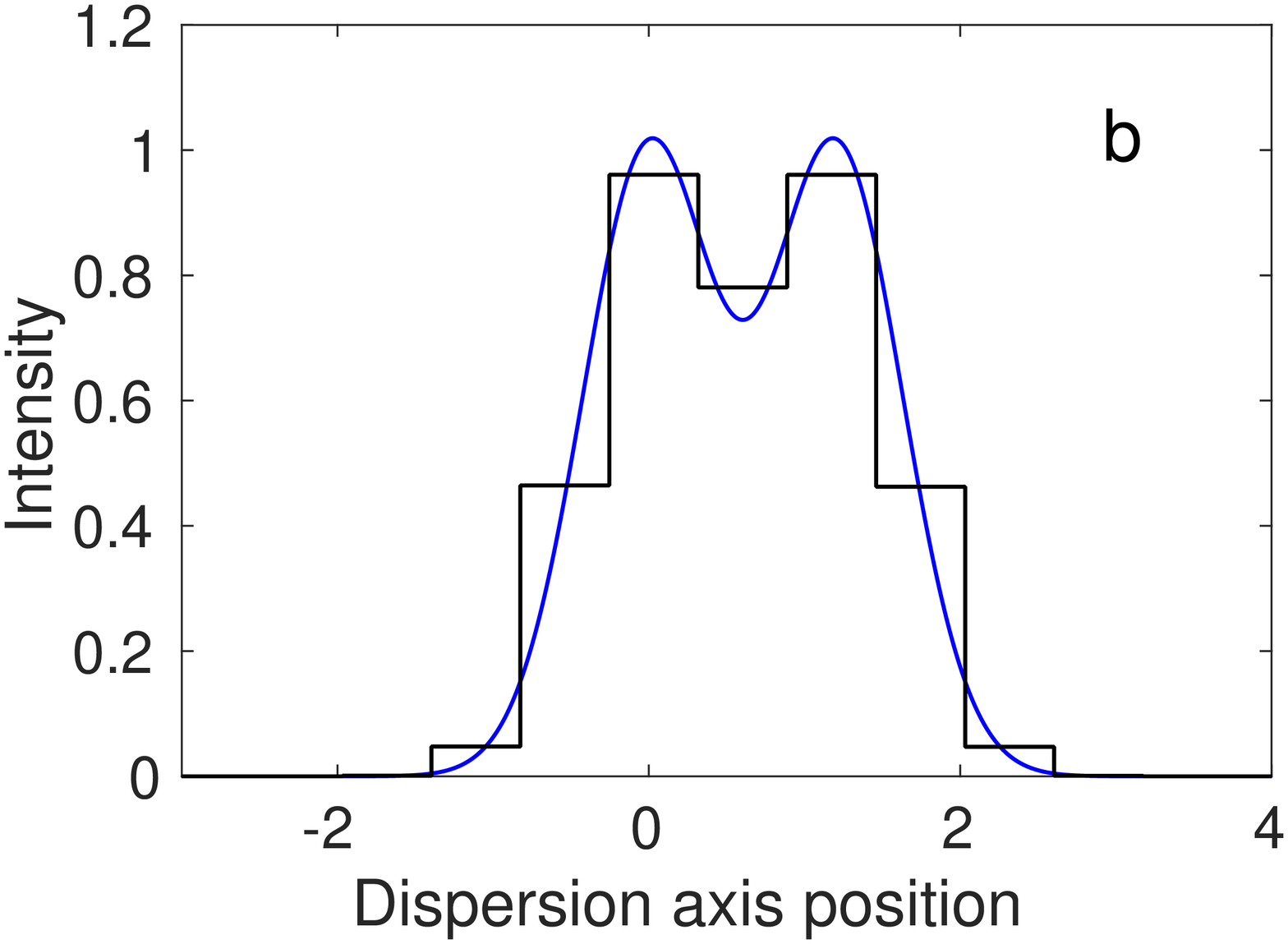}
\includegraphics[angle=0, scale=0.22]{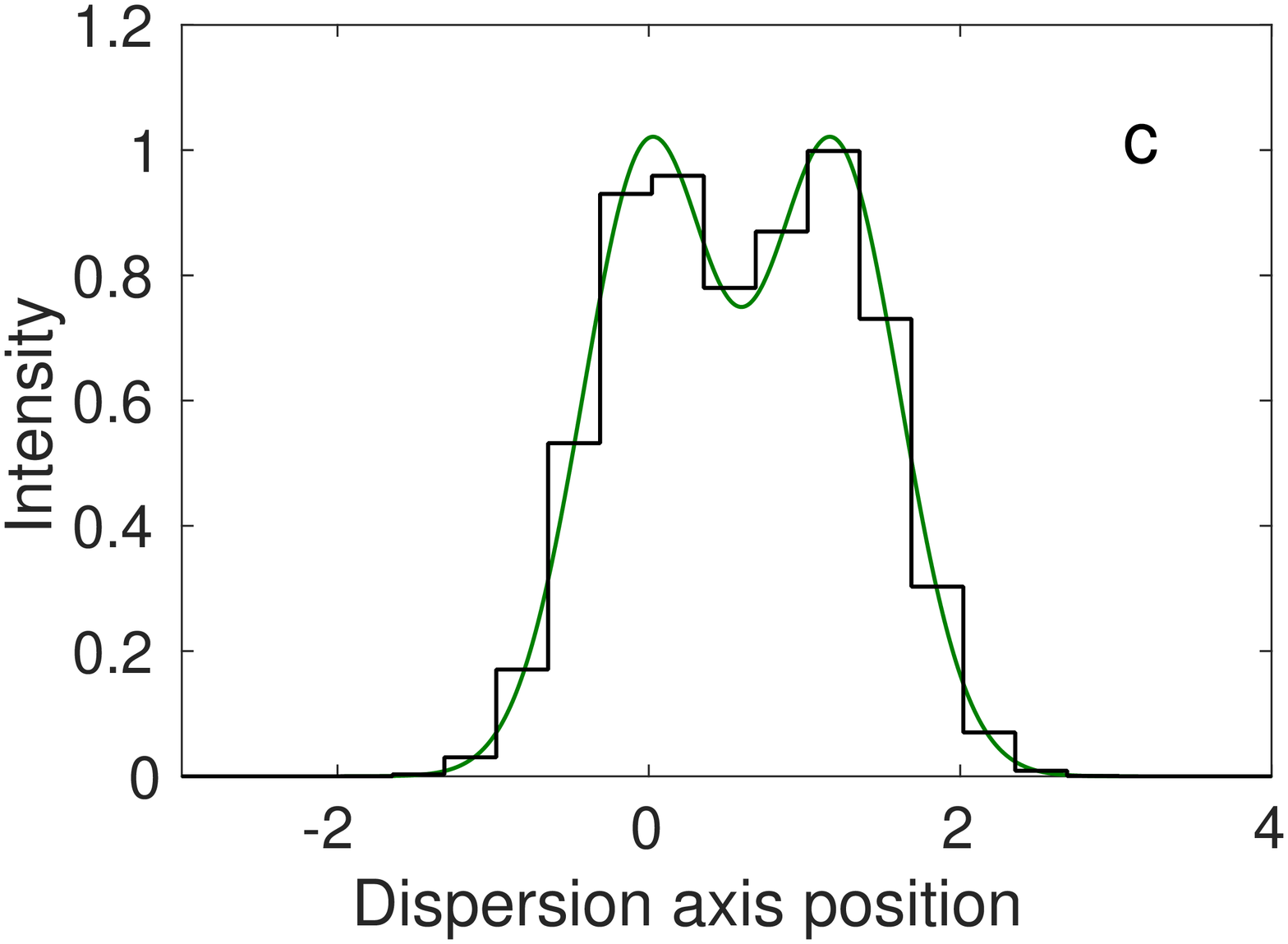}
\caption{LSF plots for the three cases indicated in Figure \ref{fig:Fig_6_164}}
\label{fig:Fig_6_65a}
\end{center}
\end{figure*}

Figure \ref{fig:Fig_6_65a} shows the intrinsic and pixellated LSFs for the three cases pointed
out in Figure \ref{fig:Fig_6_164}. It is clear why a larger separation is needed in case `a'
than in case `b' to maintain the 81.1\% relative minimum. The structure of the minimum
separation curves is one illustration of the complex non-linear effects which are introduced by
sampling. Even at 2 pixels/FWHM there is a substantial (10.1\%) variation of the critical
separation with pixel phase, which may be compared with the far smaller (0.6\%) range for
the resolving power based on wavelength accuracy for the same LSF and sampling frequency as
shown in Figure \ref{fig:Fig_7_23_k}. This is due to the different resolution criteria employed
in the two cases.

A corresponding calculation (not shown) was carried out for the convolved projected circle
LSF. It gave results that are qualitatively similar to those of the Gaussian in Figures
\ref{fig:Fig_6_164} and \ref{fig:Fig_6_65a} although with smaller critical separations as a
multiple of the FWHM.

\section{THE FOURIER VIEW OF SAMPLING} 
\label{sec:fourier}
\subsection{Overview}
\label{sec:fourier overview}
The performance of optical imaging systems is often quantified using the Optical Transfer
Function (OTF), which is the normalised Fourier Transform of the Point Spread Function
(PSF). The modulus of the OTF is known as the Modulation Transfer Function (MTF) and gives a
direct measure of a system's ability to transmit fine detail from object to image. As well as
giving an insight into performance of a subsystem, the MTF has the useful property that for
successive components of an imaging system ({\it e.g.} atmosphere, optics, detector) the final
MTF is the product of the component MTFs, provided that the transfers between the subsystems are
incoherent. 

The process of sampling of a function on a regularly-spaced grid naturally lends itself to a
study through the use of Fourier methods, since there is a maximum spatial frequency to which
the sampling can respond.  It is well-known that of the Fourier components making up a function
(here a 1-dimensional spectrum), any with spatial frequencies higher than the Nyquist limit of
2 samples per cycle will be aliased and thus incorrectly ascribed to a lower frequency. This
section will examine the implications.  An important simplification in this analysis is that to
the extent that the LSF is constant over a region of a spectrum, that spectrum can be regarded
as the convolution of the intrinsic spectrum with the LSF (with noise added after
convolution). Thus it will suffice to examine the Fourier properties of the LSF instead of full
spectra.

\begin{figure}[h]
\begin{center}
\includegraphics[angle=0, scale=0.312]{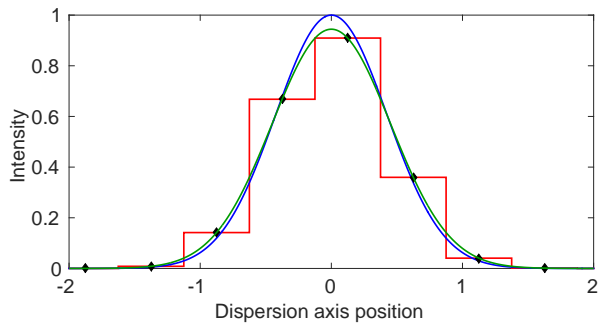}
\caption{Illustration of the effects of sampling. Blue: Gaussian intrinsic LSF, with FWHM = 1
  and peak = 1; red histogram-style line: the LSF sampled at 2 pixels/FWHM and pixel phase =
  0.25; black diamonds: the sampled points; green: the intrinsic LSF convolved with the pixel
  rectangle.}
\label{fig:Fig_6_98}
\end{center}
\end{figure}

Figure \ref{fig:Fig_6_98} illustrates the process by which a continuous LSF is rendered as a
sequence of separated samples. It also shows the broadening effect of pixel convolution - in
that the green line showing the intrinsic LSF convolved with the pixel rectangle passes
exactly through the black sample points. Thus instead of summing the LSF across the width of
each pixel, an exactly equivalent process is to convolve the LSF with the pixel rectangle and
then point sample it at the centres of the nominal pixels. In the illustrated 
case at a sampling frequency of 2 pixels/FWHM, the convolved LSF has a FWHM 1.060 $\times$
greater than the intrinsic LSF, due to the effects of pixel smoothing.

A notable complication in using the MTF to study sampling is that an assumption underpinning
the use of the MTF is that the process (in the present case, the formation of the observed
spectrum from convolution of the ideal spectrum with the LSF) should be linear and
shift-invariant. But the form of the sampled LSF depends on the pixel phase, so the process is
not shift invariant. It is nevertheless still possible to derive some useful insight from
Fourier methods. In particular, as pointed out by {\it e.g.}  Hamming (1983), {\it sin} and
{\it cos} are the eigenfunctions of equally spaced sampling.  A sinusoidal function will be
rendered with the correct functional form, amplitude and phase provided the sampling frequency
is above the Nyquist limit of 2 samples per cycle. Since the MTF treatment considers an LSF as
made up of sinusoidal Fourier components, sampling is expected to cause no errors for
components below the Nyquist frequency, provided that they have not been corrupted by aliased
components from above the Nyquist frequency.  The results in the present work support that
conclusion.

\subsection{Fourier Transforms of LSFs}
\label{sec:fourier transforms}

\begin{figure}[h]
\begin{center}
\includegraphics[angle=0, scale=0.338]{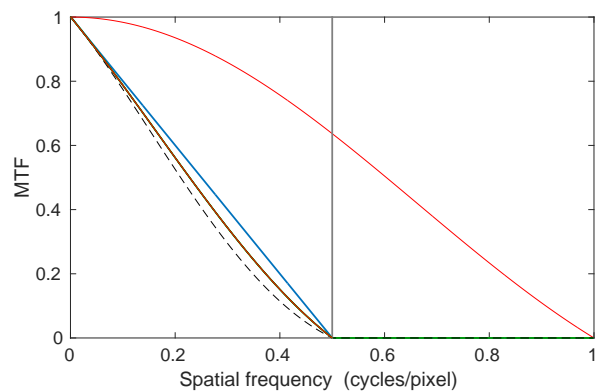}
\caption{Modulation Transfer Function of a sinc$^2$ LSF, sampled at 1.7718 pixels/FWHM. In
  order to obtain good accuracy of the transform, the sinc$^2$ subsidiary lobes were included
  out to $\pm 100 \times$FWHM. The first set of computations used 4096 points over this range,
  giving a well-sampled transform; the horizontal axis was then rescaled to show the results
  for a sampling frequency of 1.7718 pixels/FWHM. This is the reason that frequencies above the
  Nyquist frequency of 0.5 cycles/pixel can be shown. Red curve: sinc function due to smoothing
  by contiguous pixels of uniform sensitivity; blue straight line: the transform of the
  sinc$^2$ LSF; brown line: product of the above two, showing the MTF of the sampled LSF; grey
  vertical line: the Nyquist frequency for sampling; black dashed line: the LSF transform
  multiplied by two sinc factors (see text). Not visible in the plot are 6
  additional lines, all coincident with the {\it brown} line. They were computed by actually sampling
  the sinc$^2$ LSF at 1.7718 pixels/FWHM and 6 different pixel phases, and then Fourier
  Transforming. }
\label{fig:Fig_6_83_1}
\end{center}
\end{figure}

The first case considered, shown in Figure \ref{fig:Fig_6_83_1}, is for a sinc$^2$ LSF. It is
well known that the Fourier Transform of a sinc function is a rectangle, while that of a
sinc$^2$ function is a triangle peaked at the origin ({\it e.g.} Bracewell 1978). The FWHM of
\begin{equation}
{\rm sinc}^2(x) = \bigg ( \frac{\sin(\pi x)}{\pi x}\bigg )^2
\label{eqn:sinc^2}
\end{equation}
is 0.88589, and its Fourier Transform is band-limited to frequencies less than 1. Hence a
sinc$^2$ function with FWHM = 1.77178 pixels has a transform band-limited to frequencies less
than 0.5 cycles/pixel, {\it i.e.} the Nyquist frequency for sampling. In Figure
\ref{fig:Fig_6_83_1} the straight blue line shows the normalised Fourier Transform of sinc$^2$,
when sampled at the Nyquist frequency. The red curve which reaches zero at twice the Nyquist
frequency is a sinc function which represents the effect of pixel convolution on the spatial
frequencies (since sinc is the Fourier Transform of the rectangle representing uniform pixel
sensitivity). It reaches a null at one cycle per pixel, where pixel smoothing would result in
zero modulation transfer. The brown curve shows the product of these two functions, {\it i.e.}
the MTF of a sinc$^2$ LSF sampled by contiguous rectangular pixels of uniform sensitivity, at
1.7718 samples/FWHM. (These three curves were computed using fine sampling and then rescaling
the horizontal axis.)

The dashed curve relates to the statement ({\it e.g.}  Boreman (2001), Fischer et al. (2008)),
that in addition to the sinc factor due to convolution with the rectangular pixels, there is
another identical sinc factor from the sampling process itself, due to `sample scene phase
averaging'. But after convolving the LSF with the pixel rectangle, the sampling is then done at
{\it points} representing the centres of pixels. Thus there is no additional convolution with a
rectangle, and no further sinc factor in the MTF. This is demonstrated in Figure
\ref{fig:Fig_6_83_1} because the brown curve also includes 6 Fourier Transforms of the LSF
actually sampled at 1.7718 samples/FWHM. There is no dependence on pixel phase, and all
transforms match the expectation that one sinc factor is appropriate. This conclusion is
supported by the analysis of Yaroslavsky (2013).

\begin{figure}[h]
\begin{center}
\includegraphics[angle=0, scale=0.237]{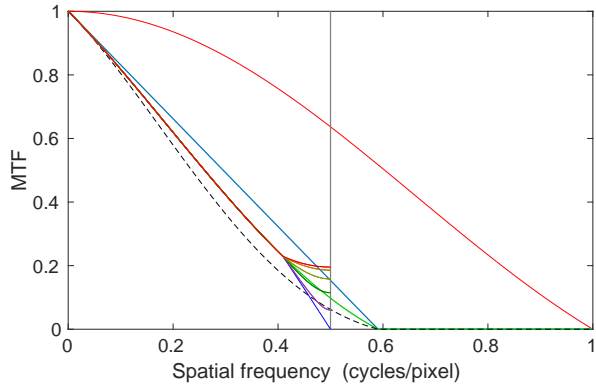}
\caption{Modulation Transfer Function of a sinc$^2$ LSF, sampled at 1.5 pixels/FWHM. Otherwise
  as for Figure \ref{fig:Fig_6_83_1}. In this case the curves for the 6 different pixel phases
  diverge sharply at spatial frequencies where aliasing occurs. (The latter curves must
  terminate at the Nyquist frequency because they were computed by actually sampling at 1.5
  pixels/FWHM.) }
\label{fig:Fig_6_83}
\end{center}
\end{figure}

To illustrate the utility of the frequency approach, Figure \ref{fig:Fig_6_83} shows again the
MTF of a sinc$^2$ LSF, but this time undersampled at 1.5 pixels/FWHM. The maximum spatial
frequency of the LSF is now 0.5906 cycles/pixel, so frequencies between 0.5 and 0.5906 are
aliased, with their Fourier components folded back into the range 0.4094 - 0.5. It is precisely
in this range that the 6 curves for different pixel phases diverge from the mean. This
demonstrates that pixel phase dependence occurs only for spatial frequencies corrupted by
aliased signal: for uncorrupted frequencies the amplitude of a Fourier component is 
correctly measured irrespective of  pixel phase, so the MTF, which measures the relative amplitude of
components, is unaffected.  Figure \ref{fig:Fig_7_23_c} shows that the same applies for
$\sigma_{\lambda}$ as a function of sampling frequency - all pixel phases give the same result
until aliasing begins at low sample rates. This behaviour is as expected since sinusoids are
the eigenfunctions of equally spaced sampling.

\begin{figure}[h]
\begin{center}
\includegraphics[angle=0, scale=0.438]{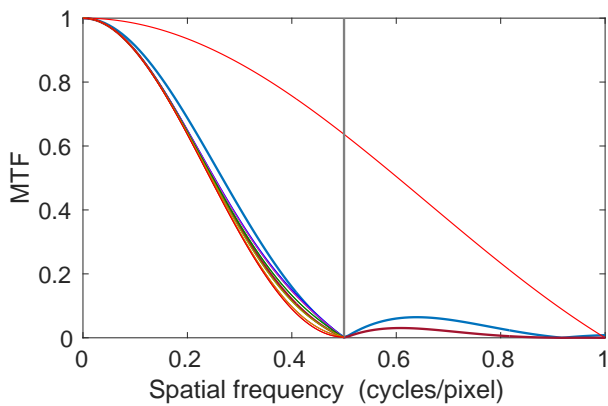}
\caption{Modulation Transfer Function of the convolved projected circle LSF, sampled at 2
  pixels/FWHM.  The colour coding of curves is as for Figures \ref{fig:Fig_6_83_1} and
  \ref{fig:Fig_6_83}. For this sampling frequency and LSF shape there is a null at 0.498,
  {\it i.e.} near the Nyquist frequency.}
\label{fig:Fig_6_88_1}
\end{center}
\end{figure}

\begin{figure}[h]
\begin{center}
\includegraphics[angle=0, scale=0.438]{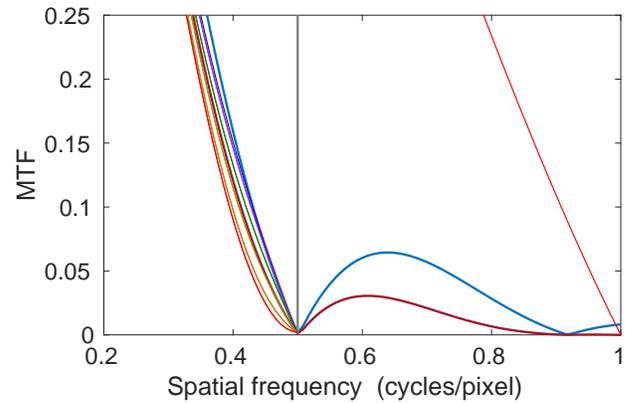}
\caption{Enlarged view of the lower part of Figure \ref{fig:Fig_6_88_1}, showing more clearly
  the curves for 6 different pixel phases. }
\label{fig:Fig_6_88_2}
\end{center}
\end{figure}

The above discussion of the sampling MTF for a sinc$^2$ LSF has illustrated important points,
but in order for the sinc$^2$ function to properly exhibit its band-limited nature, a large
number of subsidiary lobes were included in the evaluations. This is quite unrealistic for
astronomical spectra, where noise and complex structure limit the effective LSF to the main
lobe only. Thus as a final example, the MTF of the convolved projected circle LSF is shown in
Figures \ref{fig:Fig_6_88_1} and \ref{fig:Fig_6_88_2}. As expected, the steep sides of this LSF
(even after convolution with a Gaussian) result in substantial high spatial frequency
components and at the sampling frequency of 2 pixels/FWHM there is considerable amplitude
beyond the Nyquist limit. The aliasing of these components results in the curves for the 6
different pixel phases being distinct over most of the observable spatial frequency
range. This is consistent with the $\sigma_\lambda$ curves of Figure
\ref{fig:projcircleconv_sigma_lambda} which show pixel-phase dependence over essentially the
entire sample frequency range plotted.

\section{MISCELLANEOUS COMPLICATIONS} 
\label{sec:complications}

\subsection{Non-uniform pixel sensitivity}
\label{sec:non-uniform}
The results described above assumed that each pixel has uniform sensitivity which drops sharply
to zero at the pixel boundary, {\it i.e.} the pixel response as a function of position is a
rectangular function. This ideal shape will not be achieved in practice - the response may drop
off towards the edges of the pixel, possibly in an asymmetric fashion, and in thick CCD chips
there may be charge diffusion which leads to some response to photons which actually arrived in
adjacent pixels (Widenhorn et al.2010).  Jorden et al.(1994) used a small-diameter light spot
to measure the intrapixel sensitivity variations of a number of CCDs and found variations of
order 10\%.  Barron et al.(2007) used a similar technique to measure sub-pixel sensitivity
variations in a number of near-IR detectors and found high QE detectors had less than 2\%
variation, while moderate QE devices showed strong asymmetric intrapixel structure. Departures
from ideal performance of thick chips were discussed by Stubbs (2014).

Lauer (1999b) examined this issue for the case of undersampled HST images, and showed how to
construct maps of intrapixel sensitivity variations. His analysis used a set of dithered images
which allow the effective PSF ({\it i.e.} the optical PSF convolved with the pixel response
function) to be obtained with fine sampling. (This is essentially the same process as used here
to create the AAOmega LSF in Figure \ref{fig:Fig_7_23_e}.)

Intrapixel variations can be expected to cause minor departures of the actual
$\sigma_{\lambda}$ curves from those illustrated in Figures \ref{fig:Fig_7_23_a} to
\ref{fig:AAOmega_sigma_lambda}, which are in any case intended only as illustrations of some
possible forms. More important in practice is the accurate determination of the effective
LSF. Sensitivity loss at the edge of each pixel and/or leakage from adjacent pixels would
smooth out the abrupt change of signal from one pixel to the next as the illumination peak
moves - hence the `quad-cell' effect which resulted in low $\sigma_{\lambda}$ for pixel phase
0.5 in Figures \ref{fig:Fig_7_23_a}, \ref{fig:Fig_7_23_b} and \ref{fig:Fig_7_23_c} would be
diminished and the local maximum in $\sigma_{\lambda}$ may not occur ({\it e.g.} Spinhirne et
al 1998). Note that the `effective LSF', {\it i.e.} the instrumental LSF convolved with the
pixel response as found in Figure \ref{fig:Fig_7_23_e}, automatically includes the effects of
any departure of the pixel response from the ideal rectangular function $-$ without those
departures being known explicitly.

Intrapixel sensitivity variations can be expected to have a significant effect on position
bias errors, if they interact with high spatial frequency substructure of the optical LSF. For
any given system detailed measurements would be needed to quantify the effects.

A more subtle issue is the non-linear response known as the `brighter-fatter' effect, whereby
accumulated charge affects the apparent width of pixels, leading to bright objects appearing to
be up to a few percent wider than faint objects (Antilogus 2014; Rasmussen 2014; Guyonnet
2015). However, the pixel sampling frequency is not of particular importance in that case.

\subsection{Dithering and non-contiguous pixels}
\label{sec:dithering}   
While detectors with contiguous pixels represent the most common case, it is also appropriate
to examine what happens when the pixel spacing is not equal to the pixel width. The spacing is
less than the width when dithered (sub-stepped) exposures are combined, while the spacing would
be greater than the width for sparse arrays. In all cases, the process can be considered as
convolution of the intrinsic LSF with the pixel response ({\it i.e.} convolution with the pixel
{\it width}) and then point sampling at the appropriate spacing, which is less than the width
for dithering, equal to the width for contiguous pixels, or greater than the width for sparse
arrays. The use of dithering in improving undersampled images, particularly those from HST, has
been studied by many authors, {\it e.g.} Lauer (1999a; 1999b), Bernstein (2002), Fruchter and
Hook (2002), Fruchter (2011) and Boreman (2001). Lauer noted that for accurate reconstruction
of a sampled function, it must be band-limited, {\it i.e.} it should avoid corruption of
Fourier components by aliasing.

\begin{figure}[h]
\begin{center}
\includegraphics[angle=0, scale=0.34]{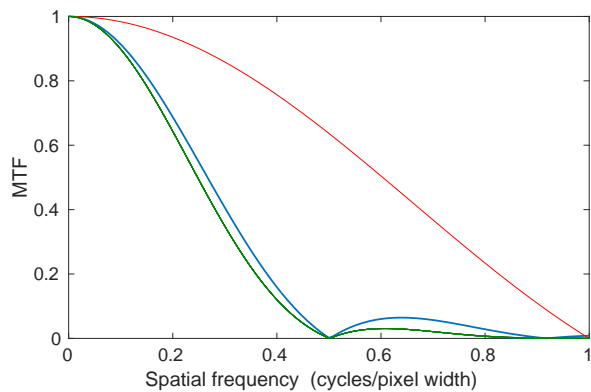}
\caption{Modulation Transfer Function of the convolved projected circle LSF, sampled at 2 pixel
  widths per FWHM and dithering with pixel spacing = 0.5 $\times$ pixel width.  The colour
  coding of curves is as for Figures \ref{fig:Fig_6_83_1} and \ref{fig:Fig_6_83}. The green line
shows 6 coincident lines for different pixel phases. }
\label{fig:Fig_6_112}
\end{center}
\end{figure}

The MTF is useful in showing the effects of varying the sampling rate in this way. For example,
Figure \ref{fig:Fig_6_112} shows the MTF of the convolved projected circle LSF sampled at 2 pixel widths
per FWHM as in Figure \ref{fig:Fig_6_88_1}, but with the data combined with a second exposure
offset by half a pixel width, so that the final pixel spacing is half of the pixel width. This
results in 4 pixel spacings per FWHM.  The horizontal axis shows spatial frequencies from 0 to
1 cycle per pixel width; with dithering this is equivalent to 0 to 0.5 cycles per pixel
spacing. Since the Nyquist frequency is related to sampling, not pixel convolution, the entire
range shown in Figure \ref{fig:Fig_6_112} is below the Nyquist frequency. Comparison of the two
Figures shows the benefit of dithering: the second lobe of the LSFs MTF, which lay beyond the
Nyquist frequency without dithering, is now correctly sampled. As a result, there is negligible
corruption due to aliasing and negligible dependence on pixel phase. This is an example of an
LSF which has significant high spatial frequency content, and benefited from sub-stepping to
reduce the pixel spacing.

\subsection{Poisson noise}
\label{sec:poisson}
The above examinations of systematic bias errors, separation of closely spaced features and the
Fourier picture are independent of the noise characteristics of the data, but the treatment of
wavelength, width and peak errors in Sections \ref{sec:wavelength accuracy}, \ref{sec:width
  accuracy} and \ref{sec:peak accuracy} and the consistent resolving power scale in Section
\ref{sec:consistent scale} do assume noise that is constant across pixels. This is suitable for
weak emission or absorption lines on a spectral continuum level, or any spectrum where detector
noise dominates. However, for strong features where Poisson shot noise from the observed object
dominates, the assumption of constant noise can be only an approximation. It is therefore
appropriate to examine briefly the differences in the case of Poisson noise.

\begin{figure}[h]
\begin{center}
\includegraphics[angle=0, scale=0.34]{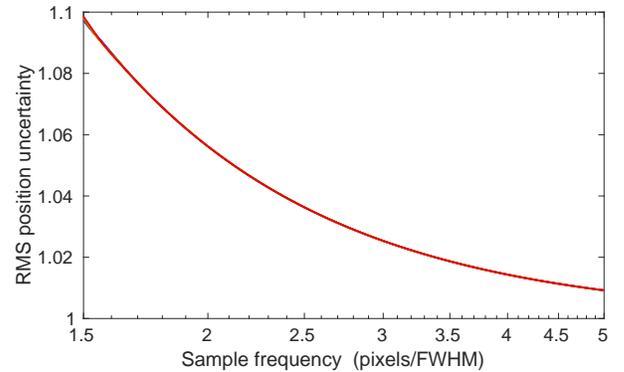}
\caption{Wavelength uncertainty vs sampling frequency for a Gaussian LSF subject to Poisson
  noise. The position (wavelength) is taken from the centroid of the observed LSF. The full range
  of pixel phases is shown as 6 curves, but they are coincident except for minor differences
  near 1.5 pixels/FWHM. The curves have been normalised to unity at very large sample frequency.}
\label{fig:Fig_6_124}
\end{center}
\end{figure}

The discussion will be limited to the case of a Poisson noise dominated unresolved emission
peak sitting on zero background level. This would describe strong emission features in a
spectrum. Since the Poisson noise approaches zero as the LSF intensity drops in the wings of
the profile, there is no need to make a least-squares fit of the LSF to the data in order to
find the location (wavelength). Instead, the centroid can be used directly. In the limit of
fine sampling of a Gaussian peak the RMS centroid error is given by the simple formula:
\begin{equation}
\sigma_{\lambda} = \frac{\sigma_{\rm G}}{\surd N }
\label{eqn:Poisson simple}
\end{equation}
where $\sigma_{\rm G}$ is the FWHM/2.3548 of the Gaussian LSF and $N$ is the total number of
photon counts in the peak. This is the same as the formula for the standard error of the mean
of a Gaussian statistical distribution. When the peak has been split into finite-width pixels,
the equation for the variance of the centroid is 
\begin{equation}
\sigma _{C_P}^2 = \frac{1}{N^2}  \sum F_i (x_i - C_P)^2.
\label{eqn:Poisson pixellated}
\end{equation}
where $F_i$ is the count in pixel $i$ located at $x_i$ and $C_P$ is the position of the
pixellated centroid.  This expression is often used in the context of the position uncertainty
of the image spots formed by a Shack-Hartmann wavefront sensor ({\it e.g.} Rousset 1999). It
has been used here to evaluate the RMS centroid uncertainty as a function of sampling
frequency, as shown in Figure \ref{fig:Fig_6_124}. This may be compared with Figure
\ref{fig:Fig_7_23_a} which shows the same for constant (normally-distributed) noise. Although
the two Figures show a broadly similar increase of wavelength uncertainty for coarser sampling,
it is notable that in the Poisson case there is much less dependence on pixel phase, with
minimal separation of curves at different pixel phase even at 1.5 pixels/FWHM. The maximum
enhancement of the RMS error over the fine-sampling limit is only $1.098 \times$, as compared
with the range $1.093 - 1.235 \times$ for constant noise.\footnote{ A Table of the factor by
  which the RMS centroid uncertainty is increased by pixellation over that given by equation
  \ref{eqn:Poisson simple} was given by Goad et al. (1986; referenced by Rousset 1999). However
  it appears that the factors given by Goad et al. are for correction of the variance not the
  RMS, since their square roots agree with the values in Figure \ref{fig:Fig_6_124} (which has
  been verified by Monte Carlo simulations).}

Finding the parameters of such a pixellated Poisson-dominated peak is equivalent to fitting a histogram
from a counting experiment in the particle physics context, and formulas for the bias, variance
and skewness of the distribution of any parameter derived from such  data were given by Eadie
et al.(1971).

\section{CONCLUSIONS} 
\label{sec:conclusions}

There are a number of consequences  of sampling spectra into pixels, and this paper is
intended to illustrate the principal effects which are of concern to the end-users of such
sampled data. In summary:

1) Random noise errors in wavelength are increased by sampling (Section \ref{sec:wavelength
  accuracy}). Uncertainties are typically $\sim 10 - 20$\% worse at a sample frequency of 2
pixels/FWHM but depend on the functional form of the Line Spread Function (LSF). The important
case of the projected circle convolved with a Gaussian (representing a projected multi-mode
fibre with some spectrograph aberrations) shows strong dependence on the pixel phase ({\it
  i.e.} the position of a spectral feature with respect to the pixel centre), with
uncertainties increased by 20\% at 2 pixels/FWHM and certain pixel phases.

2) If the width of a spectral feature is to be determined, the effects of pixellation on random
noise errors are considerably more severe than for wavelength (Section \ref{sec:width
  accuracy}), especially below 2 pixels/FWHM, and they are strongly dependent on pixel phase.

3) Pixellation causes only a minor increase in the random noise of the fitted peak amplitude
of an unresolved spectral feature (Section \ref{sec:peak accuracy}). Increases of 5\% or less
at 2 pixels/FWHM are to be expected, but with some dependence on pixel phase.

4) Pixellation tends to smooth out the relative minimum between two closely spaced emission
lines (or equivalently, the relative maximum between two absorption lines). If one wishes to
see a relative minimum of 81\% of the peak (equal to the Rayleigh criterion separation for two
finely-sampled sinc$^2$ LSFs) then the separation required is significantly increased by
sampling, but in a complex manner due to the effects of pixel phase (Section \ref{sec:resolving
  features}).

5) As demonstrated in Robertson (2013; Paper 1), the FWHM is a poor measure of spectral
resolution, when LSFs of different form need to be compared. In Section \ref{sec:consistent
  scale} the method for calculating resolving power on a consistent scale based on wavelength
accuracy is extended from that given in Paper 1 to include effects of pixellation.

6) Pixellation of spectra can produce systematic bias errors in wavelength that depend on pixel
phase and the method of wavelength determination, but are not reduced by high signal/noise
data. As shown in Section \ref{sec:bias} such errors may be negligible for well-sampled
symmetric LSFs. But they are greatly increased by asymmetry in the LSF and/or high spatial
frequency components that are not adequately sampled by the detector. Thus any fine structure
in the intrinsic LSF ({\it i.e.} the unsampled image as it falls on the detector plane) will
have a significant role in producing wavelength bias errors, and a smooth symmetrical PSF that
varies at most slowly across the detector is desirable for high precision work.

7) The Modulation Transfer Function (MTF; normalised amplitude of the Fourier Transform of the
LSF) can show the extent to which spatial frequencies making up the LSF are aliased due to
inadequate sampling frequency. Any Fourier component which is aliased will then corrupt another
component which is below the Nyquist frequency and would have otherwise been correctly
recorded. Pixel-phase dependence develops for spatial frequencies that have been corrupted by
aliased signal from above the Nyquist frequency.

8) There has developed in the literature a practice of referring to a sampling frequency of 2
samples per FWHM as being the Nyquist limit. This is incorrect, since it is not the same as 2
samples per cycle of a sinusoid, and most common LSFs have aliased high frequency components
when sampled at 2 pixels/FWHM. Nevertheless, it is true that for most LSF forms resulting from
spectrographs, 2 samples/FWHM is a reasonable minimum if high precision is not required. In the
case of a diffraction-limited slit input, 2 samples/FWHM is slighly more than needed to avoid
any aliasing (but this assumes that subsidiary lobes of the diffraction pattern can be included
in the analysis - which is unlikely in spectra with noise and many features). For high
precision work with typical non-diffraction-limited LSFs, a larger sampling frequency should be
used. For example Chance et al.(2005) recommend 4.5 $-$ 6.5 pixels/FWHM of a Gaussian LSF
after pixel convolution to avoid any significant aliasing, and the HARPS planet-finder
spectrograph (Mayor et al, 2003) uses 3.2 pixels/FWHM.

9) A recommendation from this work is that it is desirable for designers of spectrographic
instruments to carry out simulations using computed LSFs (suitably smoothed to remove spurious
high frequency features such as from a finite number of rays traced, but retaining all `real'
fine structure). Then the extent to which noise in parameter estimates is increased, the
amount of pixel-phase dependence and the bias errors can be found for any proposed camera speed
and detector pitch. 

10) For instruments where the end users have choices regarding sampling frequency ({\it e.g.}
from on-chip binning, or different spectrograph configurations) it is highly desirable that the
instrument team should carry out simulations using the actual LSF and so provide in the user
manual data regarding the effects on noise, bias, separation of peaks, pixel-phase dependence
etc. which can aid the user in deciding which configuration to use.

\section*{Acknowledgments} 
I thank Tayyaba Zafar and Sarah Brough for providing AAOmega data, Richard Hook and Will
Saunders for helpful discussions, and Andrew Sheinis for pointing out one of the references.

\vspace*{6mm}
\noindent
{\bf REFERENCES}\\

\noindent
Anderson, J. and King, I.R.  PASP 112, 1360, 2000.\\[-1mm]

\noindent
Antilogus, J., Astier, P.,  Doherty, P., Guyonnet, A, and Regnaulta, N.  Inst. 9, C03048 2014. \\[-1mm]

\noindent
Barron, N., Borysow, M., Beyerlein, K., Brown, M., Lorenzon, W., Schubnell, M., Tarle´, G.,
Tomasch, A. and Weaverdyck, C. PASP 119 466 2007. \\[-1mm]

\noindent
Bernstein, G.  PASP 114, 98, 2002. \\[-1mm]

\noindent
Bickerton, S.J. and Lupton, R.H. MNRAS 431, 1275, 2013.\\[-1mm]

\noindent
Boreman, G.D. {\it Modulation Transfer Function in Optical and Electro-Optical Systems} SPIE
Press 2001.  \\[-1mm]

\noindent
Bracewell, R.N. {\it The Fourier Transform and its Applications} McGraw-Hill 1978.\\[-1mm]

\noindent
Bracewell, R.N. {\it Two Dimensional Imaging,} Prentice Hall p365, 1995.\\[-1mm]

\noindent
Chance, K., Kurosu, T.P. and Sioris, C.E. Appl. Opt.  44, 1296  2005.\\[-1mm]

\noindent
Clarke, T.W., Frater, R.H., Large, M.I., Munro, R.E.B. and Murdoch,
H.S. Aust. J. Phys. Astrophys. Suppl. 10, 3, 1969.\\[-1mm] 

\noindent
Eadie, W.T., Dryard, D., James, F.E., Roos, M. and Sadoulet, B. {\it Statistical Methods in
  Experimental Physics} North-Holland p. 146, 1971.\\[-1mm]

\noindent
Fischer, R.E., Tadic-Galeb, B. and Yoder, P.R. {\it Optical System Design} McGraw-Hill 2008.\\[-1mm]

\noindent
Fruchter, A.S. PASP 123, 497, 2011.\\[-1mm]

\noindent
Fruchter, A.S. and Hook, R.N. PASP, 114, 144, 2002.\\[-1mm]

\noindent
Guyonnet, A., Astier, P., Antilogus, P., Regnault, N. and Doherty, P. Astron. Astrophys. 575,
A41 2015.\\[-1mm] 

\noindent
Hamming, R.W.  {\it Digital Filters} Prentice-Hall 2nd ed, p30, 1983.\\[-1mm]

\noindent
Jorden, P.R., Deltorn, J-M and Oates, A.P. Proc. SPIE 2198, 836, 1994.\\[-1mm]

\noindent
Lauer, T.R.  PASP 111, 227, 1999a.\\[-1mm]

\noindent
Lauer, T.R. PASP 111, 1434, 1999b.\\[-1mm]

\noindent
Mayor, M. et al. ESO Messenger 114, 20, 2003.\\[-1mm]

\noindent
Rasmussen, A. J. Inst 9, C04027 2014.\\[-1mm]

\noindent
Robertson, J.G. PASA 30, e048 2013 (Paper 1).\\[-1mm]

\noindent
Rousset, G. in  {\it Adaptive Optics in Astronomy} ed. Roddier, F.  Cambridge University Press p
115, 1999.\\[-1mm]

\noindent
Saunders, W., Bridges, T., Gillingham, P., Haynes, R., Smith, G.A., Whittard, J.D.,
Churilov, V., Lankshear, A., Croom, S., Jones, D. and Boshuizen, C. SPIE 5492, 389, 2004.\\[-1mm]

\noindent
Saunders, W. Proc. SPIE 9147 60 2014.\\[-1mm]

\noindent
Spinhirne, J.M., Allen, J.G., Ameer, G.A., Brown, J.M., Christou, J.C., Duncan, T.S., Eager,
R.J., Ealey, M.A., Ellerbroek, B.L., Fugate, R.Q., Jones,G.W., Kuhns, R.M., Lee, D.J., Lowrey,
W.H., Oliker, M.D., Ruane, R.E., Swindle, D.W., Voas, J.K., Wild, W.J., Wilson, K.B. and John
L. Wynia, J.L. Proc. SPIE  3353,  22,  1998.\\[-1mm]

\noindent
Stubbs, C.W.  J. Inst. 9, C03032, 2014.\\[-1mm]

\noindent
Vollmerhausen, R.H., Reago, D.A. and Driggers, R.G. {\it Analysis and Evaluation of Sampled
  Imaging Systems} SPIE Press 2010.\\[-1mm]

\noindent
Widenhorn, R., Alexander Weber-Bargioni, A., Blouke, M.M., Bae, A.J. and Bodegom, E.
Opt. Eng. 49, 044401, 2010.\\[-1mm]

\noindent
Yaroslavsky, L.P. {\it Theoretical Foundations of Digital Imaging using MATLAB} CRC Press p83
2013.\\[-1mm]

\end{document}